\documentclass[aps,twocolumn,superscriptaddress,showpacs,amssymb,prx,floatfix,longbibliography]{revtex4-2}
\usepackage{graphicx}
\usepackage[]{hyperref}
\usepackage{amsmath}
\usepackage{amssymb}
\usepackage{bm}
\usepackage{xcolor}

\hypersetup{colorlinks=true,linkcolor=blue,citecolor=blue,urlcolor=blue,pdfpagemode=UseNone}

\newcommand{\slrrtext}  {spin-lattice-relaxation rate}
\newcommand{\slrr}      {$T_1^{-1}$}
\newcommand{\tmvo}      {TmVO$_4$}

\begin{document}

\title{Second order Zeeman interaction and ferroquadrupolar order in TmVO$_4$}

\author{I. Vinograd}
\affiliation{Department of Physics and Astronomy, University of California Davis, Davis, CA }
\author{K. R. Shirer}
\affiliation{Max Planck Institute for Chemical Physics of Solids, D-01187 Dresden, Germany}
\author{P. Massat}
\affiliation{Geballe Laboratory for Advanced Materials and Department of Applied Physics, Stanford University,  CA 94305, USA}
\author{Z. Wang}
\affiliation{Department of Physics and Astronomy, University of California Davis, Davis, CA }
\author{T. Kissikov}
\affiliation{Department of Physics and Astronomy, University of California Davis, Davis, CA }
\author{D. Garcia}
\affiliation{Department of Physics and Astronomy, University of California Davis, Davis, CA }
\author{M.~D.~Bachmann}
\affiliation{Geballe Laboratory for Advanced Materials and Department of Applied Physics, Stanford University,  CA 94305, USA}
\author{M. Horvati\'c}
\affiliation{Laboratoire National des Champs Magn\'etiques Intenses, LNCMI-CNRS (UPR3228), EMFL, Universit\'e
Grenoble Alpes, UPS and INSA Toulouse, Grenoble, France}
\author{I. R. Fisher}
\affiliation{Geballe Laboratory for Advanced Materials and Department of Applied Physics, Stanford University,  CA 94305, USA}
\author{N. J. Curro}
\affiliation{Department of Physics and Astronomy, University of California Davis, Davis, CA }

\date{\today}


\begin{abstract}
TmVO$_{4}$ exhibits ferroquadrupolar order of the Tm 4f electronic orbitals at low temperatures, and is a model system for Ising nematicity that can be tuned continuously to a quantum phase transition via magnetic fields along the $c$-axis.  Here we present $^{51}$V nuclear magnetic resonance data in magnetic fields perpendicular to the $c$-axis in a  single crystal that has been carefully cut by a plasma focused ion beam to an ellipsoidal shape to minimize the inhomogeneity of the internal demagnetization field. The resulting dramatic increase in spectral resolution enabled us to resolve the anisotropy of the electric field gradient and to measure the magnetic and quadrupolar relaxation channels separately. Perpendicular magnetic fields nominally do not couple to the low energy degrees of freedom, but we find a significant nonlinear contribution for sufficiently large fields that give rise to a rich phase diagram. The in-plane magnetic  field can act either as an effective transverse or longitudinal  field to the Ising nematic order, depending on the orientation relative to the principle axes of the quadrupole order, and leads to a marked in-plane anisotropy in both relaxation channels. We find that the small in-plane transverse fields initially enhance the ferroquadrupolar ordering temperature but eventually suppress the long-range order.  We tentatively ascribe this behavior to the competing effects of field-induced mixing of higher energy crystal field states and the destabilizing effects of field-induced quantum fluctuations.

\end{abstract}
\maketitle

\section{Introduction}

Electronic nematic order refers to a state in which low energy electronic degrees of freedom drive a crystal to spontaneously break discrete rotational symmetry without simultaneously breaking translational symmetry
\cite{FradkinNematicReview}. Much of the reason for the current interest in such states derives from observations that unconventional superconductivity
tends to emerge in materials exhibiting competing
ground states with different broken symmetries, including nematic,
charge, and/or spin density wave fluctuations
\cite{Kivelson1998,Vojta2009,FernandesSchmalianNatPhys2014,Baek2015,Comin2016}.
{Nuclear Magnetic Resonance (NMR) is a technique that can probe these forms of symmetry breaking locally through the magnetic and electronic charge environment to which the NMR active nucleus couples through its spin and quadrupole moment, respectively. Quadrupolar fluctuations are a measure of the nematic susceptibility and contribute to the spin-lattice relaxation rate \slrr, but have to be neglected usually when magnetic fluctuations dominate the relaxation. Recovering the quadrupolar relaxation is of particular interest in measurements of the iron-based superconductors, as their} putative nematic quantum critical point appears to be correlated with optimal superconductivity and non-Fermi-liquid behavior in the normal state, indicating a possible role for nematic fluctuations in the pairing interaction  \cite{FisherScienceNematic2012,Kuo2015,KivelsonNematicQCP2015,ScalapinoNematicQCP2015}. However, disentangling the effects of quantum critical nematic fluctuations from other competing order parameter fluctuations presents a significant challenge.  Moreover, the emergence of superconductivity preempts the low temperature behavior where a quantum phase transition may be present.  Hence there is an interest to investigate electronic nematicity on its own, without the presence of competing phases.
{Here, we investigate a sample of \tmvo, which manifests nematicity through ferroquadrupole order and whose NMR linewidth is reduced by carefully cutting the sample to an ellipsoid using a focused ion beam (FIB). Based on this innovation we recover the quadrupole relaxation rates through relaxation measurements at all satellite peaks. A further consequence of this innovation is that we have been able to demonstrate how an in-plane magnetic field, depending on the field angle, can act as either a longitudinal or transverse field for ferroquadrupole order to selectively enhance or suppress nematicity in \tmvo.}

Ferroquadrupolar order of atomic orbitals is an important realization of electronic nematicity, in which electronic orbitals spontaneously develop quadrupole moments oriented in the same direction \cite{FisherNematicQCP,Rosenberg2019}. Coupling between the quadrupolar moments and the lattice gives rise to an effective interaction between the atomic orbitals, and leads to a cooperative Jahn-Teller distortion at a temperature, $T_Q$ \cite{Gehring1975}. In many cases ferroquadrupolar order develops in insulators, without the presence of competing broken symmetry phases. \tmvo\ is an ideal example of such a material, which has been well characterized \cite{BleaneyTmVO4review,Massat2021}. The Tm ions (4f$^{12}$ with $L=5$, $S=1$, $J=6$) in this material have partially filled 4f shells that are split by a tetragonal crystal field. The ground state doublet is well separated by a gap of $\sim 77$ K to the lowest excited state \cite{Knoll1971}. Importantly, the ground state is a non-Kramers doublet, such that the first order Zeeman interaction vanishes for in-plane fields  (i.e. $g_c\sim 10$ while $g_a = g_b = 0$). The ground state can be well-described by a pseudospin ($\tilde{S} = 1/2$), in which one component, $\tilde{S}_z$, corresponds to a magnetic dipole moment oriented along the $c$-axis, while the other two components $\tilde{S}_x$ and $\tilde{S}_y$ correspond to electric quadrupole moments with $B_{2g}$ ($xy$) and  $B_{1g}$ ($x^2-y^2$) symmetry respectively \cite{MelcherCJTEreview}. {The quadrupole moments couple bilinearly to external fields that have these symmetries. This could be strains $\varepsilon_{xx}-\varepsilon_{yy}$, $\varepsilon_{xy}$ or it could be composite magnetic field variables that transform in the same way, such as $B_x^2-B_y^2$ and $B_x B_y$, where $x$ and $y$ lie along the axes of the undistorted tetragonal unit cell. Hence, although in-plane magnetic fields do not couple linearly to the quadrupole moments, they do couple quadratically, and (appropriately constructed objects that go as the square of the fields) can act as effective longitudinal and transverse fields for a quadrupole ordered state, respectively. In \tmvo,} the quadrupole moment of the doublet couples to the lattice strain, $\varepsilon_{xy}$, and the material spontaneously undergoes a tetragonal to orthorhombic distortion with $B_{2g}$ symmetry below $T_Q = 2.15$ K with orthorhombicity $\delta\approx 0.01$. The low energy degrees of freedom can be described well by the transverse field Ising model, in which Ising interactions couple the {pseudospins which represent the quadrupolar moments} and a magnetic field along the $c$-axis couples transverse to the quadrupolar direction \cite{FisherNematicQCP}. This field enhances the fluctuations of the pseudospins and can tune the system to an Ising-nematic quantum phase transition at a critical field $B_c\approx0.5$ T \cite{Massat2021}. This material thus offers an important platform to investigate quantum critical nematic fluctuations in an insulator.

The critical behavior of the ferroquadrupolar order has been well documented for fields along $c$-direction, but little is known about the behavior for in-plane magnetic fields.  Although the in-plane Zeeman interaction vanishes to first order, the second order effect can become important for sufficiently high magnetic fields \cite{Washimiya1970}. Previous nuclear magnetic resonance (NMR) measurements of \tmvo\ in this field orientation identified a scaling between the \slrrtext\ and the shear elastic stiffness constant, $c_{66}$, suggesting that the $^{51}$V  ($I=7/2$) nuclear spins couple to the Tm orbitals through the electric field gradient (EFG), giving rise to a quadrupolar relaxation channel \cite{Wang2021}.  However, the spectra were significantly broadened by inhomogeneous demagnetization fields and the anisotropic $g$-factor of the Tm ground state doublet.  In order to better discern the spectra and relaxation mechanisms at play, we utilized a focused ion beam (FIB) to shape a single crystal of \tmvo\ in an ellipsoidal shape, with a homogeneous demagnetization field.   The enhanced spectral resolution of our FIB crystal enables detailed in-plane angular dependent studies.  We find that the EFG asymmetry parameter, $\eta$, which is a direct measure of the ferroquadrupolar order parameter, depends sensitively on the direction of the in-plane magnetic field.  We interpret this behavior as a consequence of a second order Zeeman effect on the non-Kramers doublets. An in-plane magnetic field $45^{\circ}$ to the orthorhombic distortion ($\mathbf{B} \parallel [100]_T$) acts as a transverse field to the Ising order, whereas in-plane fields oriented parallel to the distortion directions ($\mathbf{B} \parallel [110]_T$) act as longitudinal fields. Consequently the former can tune the transition to a quantum phase transition, whereas the latter will broaden the phase transition.

The enhanced spectral resolution also enables us to disentangle the magnetic and quadrupolar contributions to the V spin lattice relaxation. We find these rates exhibit different temperature dependencies, and depend not only on the orientation of the applied magnetic field in the $ab$ plane, but also vary strongly with applied field strength. Surprisingly, we find that for transverse in-plane fields,  $T_Q$ is enhanced for applied fields $0 \leq B \lesssim 5$ T. This unusual behavior may reflect the influence of the excited crystal field states. For fields greater than $\sim 5$ T, the phase transition appears to exhibit a crossover rather than suppress to zero.  We speculate that this behavior is related to the dynamics of isolated two level systems once the magnetic field is sufficiently large to split the Tm ground state doublets. These unexpected observations for in-plane fields reveal rich new physics driven by nonlinear Zeeman interactions in this model system.

\section{NMR Spectra}

\subsection{Sample Shaped by FIB \label{sec:sample}}

\begin{figure}
\includegraphics[width = \linewidth]{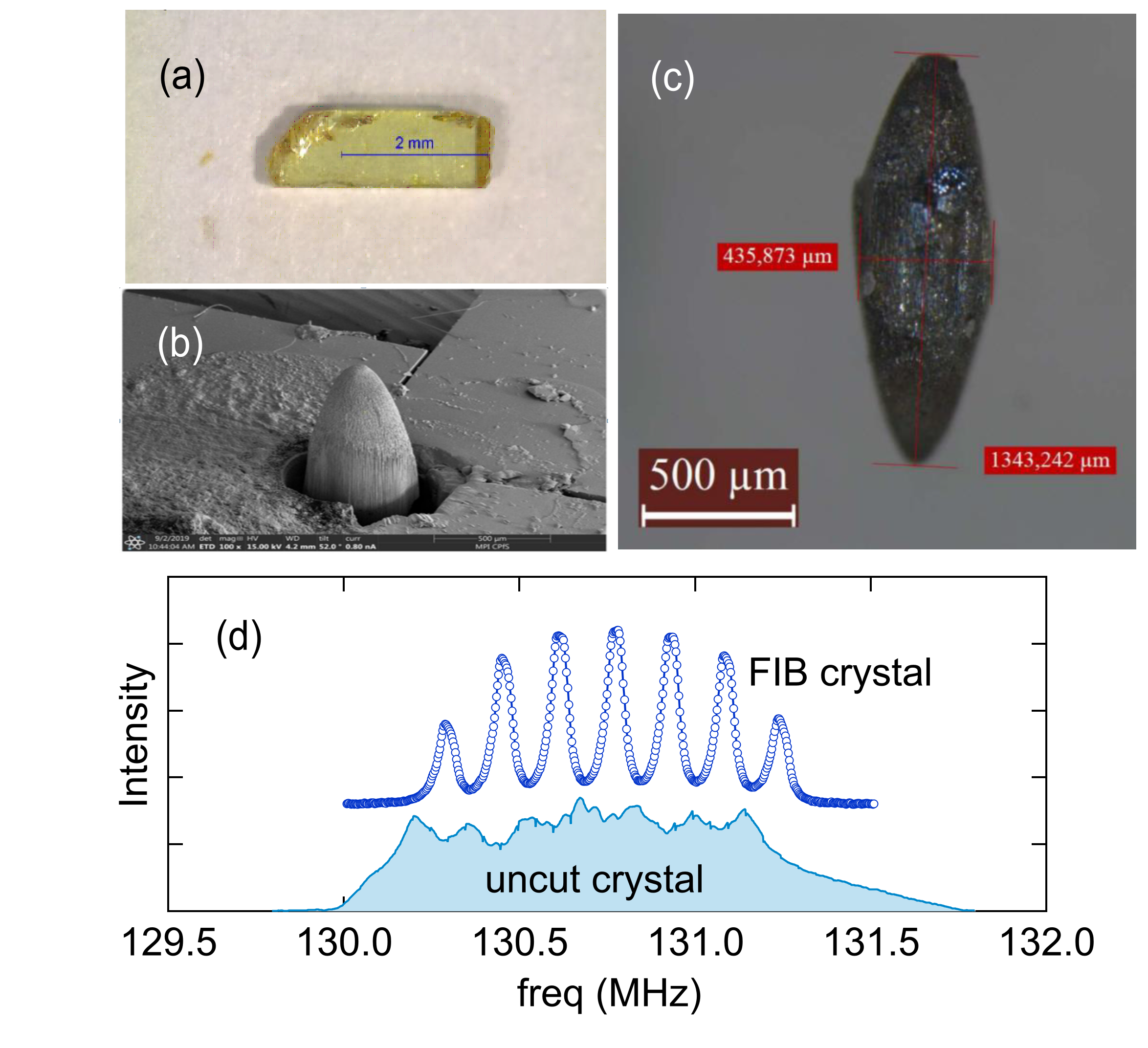}
\caption{ (a) Uncut sample of \tmvo. (b) Scanning electron microscopy (SEM) scan of the sample during the FIB process. (c) Crystal after FIB.  Al and C are deposited on the sample surface layer during the FIB processing, but do not contribute to the NMR signal. (d) $^{51}$V-NMR spectra of ellipsoidal (FIB) and uncut samples measured at $T$=10~K and $B_0$=11.7294~T, illustrating the broadening effect of the inhomogeneous demagnetization field on the seven nuclear spin transitions.
\label{fig:FIB}}
\end{figure}

Single crystals of \tmvo\ were grown from a Pb$_2$V$_2$O$_7$ flux using 4 mole percent of Tm$_2$O$_3$, following the methods described in \cite{FEIGELSON1968,Smith1974}.  As illustrated in Fig. \ref{fig:FIB}(a), the crystals grow as rectangular needle-like prisms, with the long axis along the $c$-direction. The NMR spectrum depends sensitively on the shape of the crystal and the direction of the applied magnetic field, $\mathbf{B}_0$, which polarizes the magnetic moments and creates a magnetization $\mathbf{M}$ in the sample.
Continuity of the flux-density $\mathbf{B}=\mu_0\mathbf{M}+\mathbf{B_0}$ across the sample surface requires a demagnetizing field $\mu_0\mathbf{H}_d$ in addition to $\mathbf{B}_0$ \citep{Blundell2001}. In an arbitrarily shaped sample $\mu_0\mathbf{H}_d(\bf r)$ and hence $\mathbf{B}(\mathbf{r})$ will be inhomogeneous and the NMR spectrum will be broadened because the nuclei each resonate at the local field \cite{Lawson2018}. In most cases this broadening is not sufficient to cause any significant problems for NMR, however in \tmvo\ the non-Kramers doublet has $g_c \approx 10$ and $g_{ab} =0$.  Although $\mathbf{B}_0$ can be oriented perpendicular to $c$, parts of the crystal near edges and corners tend to have components of $\mathbf{B}$ parallel to $c$, which exacerbates the broadening effect due to the large anisotropy of the susceptibility, $\chi$. 
The resulting broadening is sufficient to wash out the quadrupolar splitting ($\sim 150$ kHz) between the Tm satellites at low temperatures and high fields \cite{Wang2021}.

Fortunately, $\mu_0\mathbf{H}_d$ can be made homogeneous by cutting the sample in either a spherical or ellipsoidal shape \cite{Osborn1945}. For these studies we utilized a focused ion beam (FIB) to cut our sample to an ellipsoid with the long-axis along the $c$-axis of the crystal \citep{Moll2018}, as illustrated in Fig. \ref{fig:FIB}(b,c). The sample is cut from a carefully aligned cuboid from which calculated and programmed concentric circles are removed using by a xenon plasma FIB by Thermo Fisher Scientific with a 30~kV, 1~$\mu$A beam. Sample damage from the beam is only expected on the surface within a depth of $30-40$~nm~\citep{Eder2021,Kelley2013,Giannuzzi2011} and Energy Dispersive X-Ray Analysis (EDX) of a test surface verifies the unchanged composition of TmVO$_4$ below. The final sample diameter is 0.4~mm and the length of 1.3~mm require a total cutting time in excess of 25~h of each side. In the FIB process Al and C are deposited on the sample surface layer, however these do not affect the NMR signal from the bulk of the sample.

The cut crystal was mounted in a dual-axis goniometer, as discussed below in \ref{sec:alignment}.  Fig. \ref{fig:FIB}(d) compares the NMR spectrum of the uncut with the FIB crystal. It is clear that the magnetic broadening is dramatically reduced in the FIB crystal, such that each of the seven peaks separated by the quadrupolar splitting are clearly resolved. The ability to resolve all seven peaks is important because it enables us to extract details of the magnetic and quadrupolar contributions to the \slrrtext\ that would otherwise be inaccessible, as discussed below in Section \ref{sec:relax}. For fields $B_0 \leq 1.35$~T the broadening from demagnetizing fields is sufficiently reduced, so that a larger crystal whose sharp corners were carefully polished away by hand was utilized.

\begin{figure}
  \includegraphics[width=0.405\linewidth]{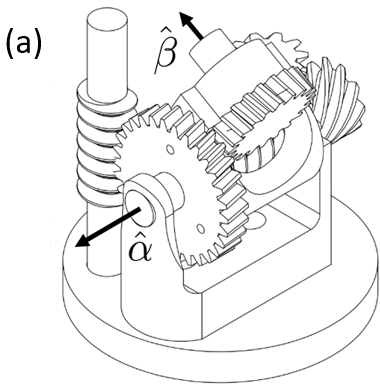}
   \includegraphics[width=0.58\linewidth]{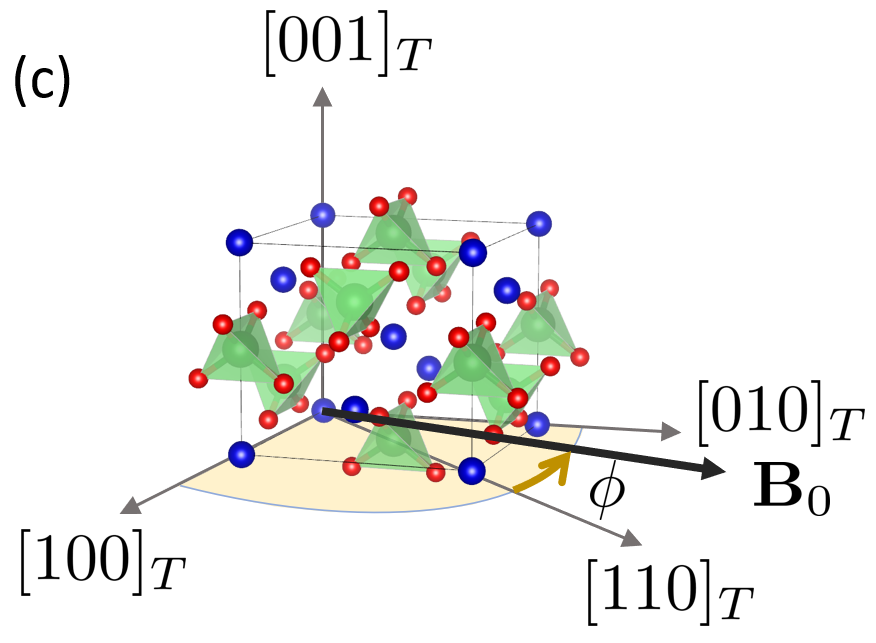}
   \includegraphics[width=\linewidth]{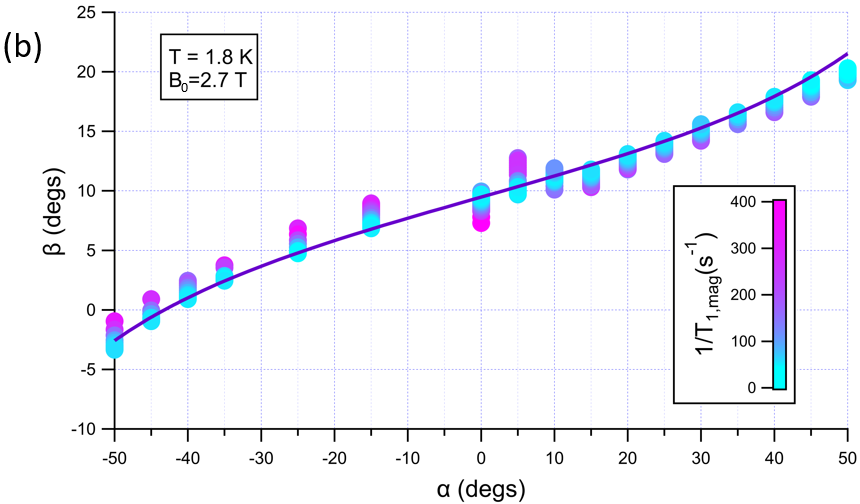}
\caption{(a) Schematic of the goniometer with rotation axes $\hat{\bm{\alpha}}$ and $\hat{\bm{\beta}}$. Note that the direction of  $\hat{\bm{\beta}}$ depends on the angle $\alpha$. (b) Measurements of the strongly anisotropic magnetic relaxation rate $1/T_{1,\rm{mag}}$ at  2.7 T  and 1.8 K. The minimal relaxation rate (light blue) corresponds to purely in-plane orientations. The functional dependence of $\beta(\alpha)$ of this minimum is given by Eq. \ref{eqn:alphabeta} and is shown as a solid line for $\gamma=13.7^{\circ}$. (c) Diagram of field orientation with respect to the tetragonal unit cell, with the angle $\phi$ being measured from the $[110]_T$ direction. Blue atoms are Tm, green are V, and red are oxygen.}
\label{fig:align}
\end{figure}

\subsection{Crystal alignment\label{sec:alignment}}

\begin{figure*}
\includegraphics[width =0.55 \linewidth]{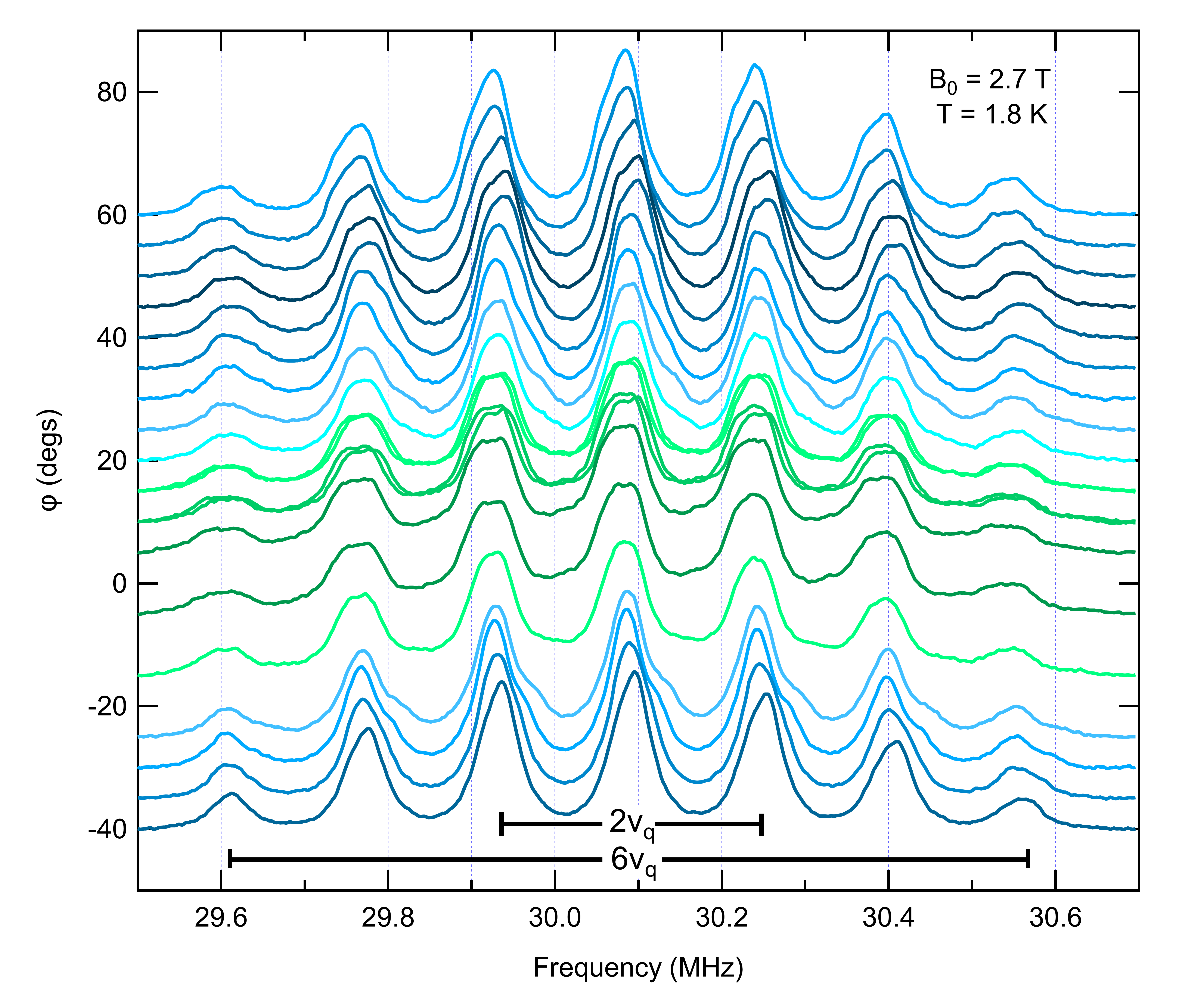}
 \includegraphics[width=0.43\linewidth]{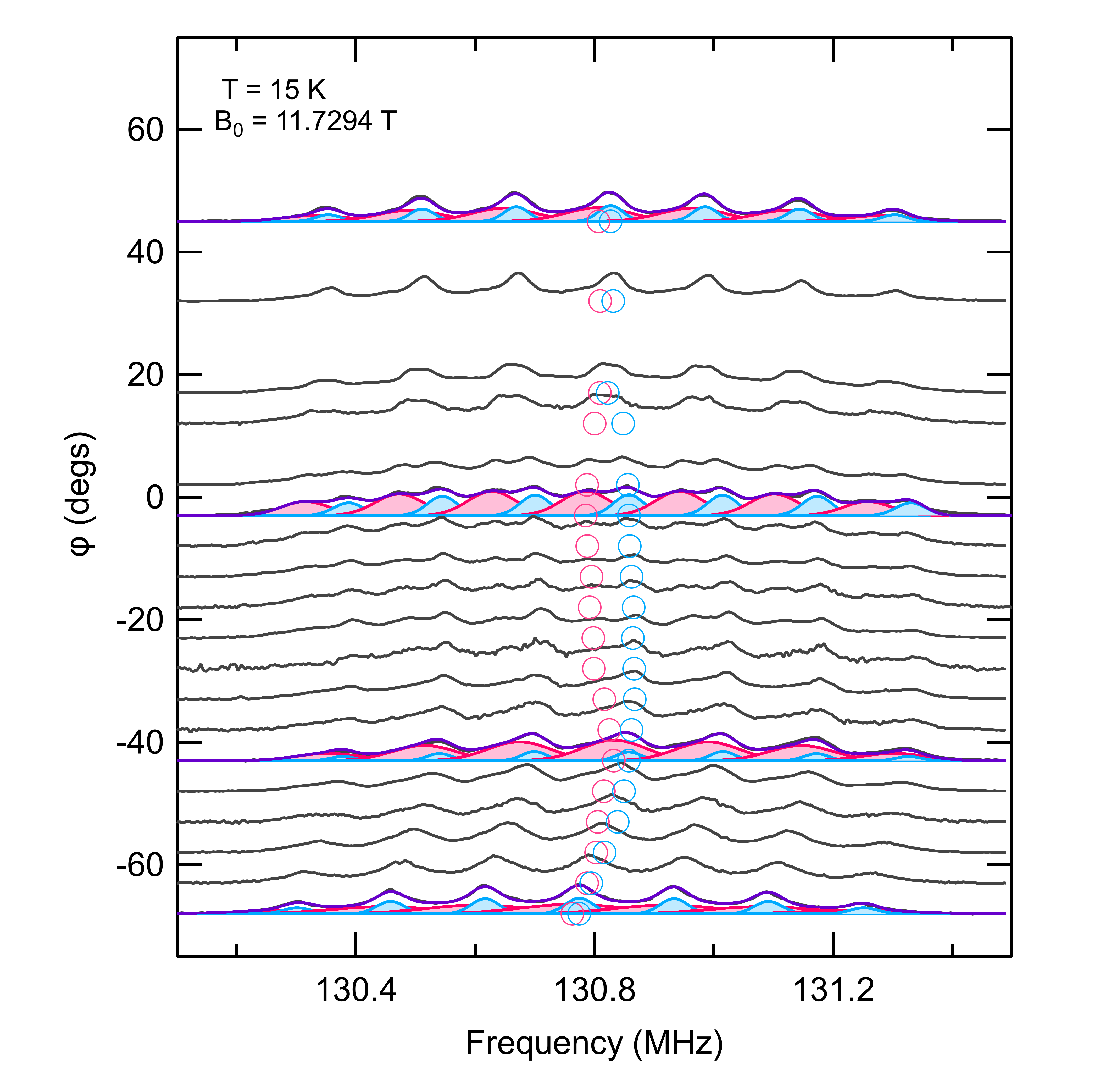}
\caption{$^{51}$V NMR spectra as a function of in-plane field angle, $\phi$, measured at (Left) 2.7 T and 1.8 K, and at (Right) $B_0=11.7294$~T and $T=15$~K. The offset corresponds to the in-plane angle $\phi$. For the high field spectra, each transition is fit with two different Gaussian peaks that have the same width independent of the transition (filled blue and red regions). The splitting between the Gaussians is clearest around $\phi=0^{\circ}$, but the splitting remains finite for most orientations due to finite asymmetry in the line shapes.The open circles correspond to the center of gravity of each set of transitions.}
\label{fig:waterfall}
\end{figure*}

In order to investigate the impact of the ferroquadrupolar ordering on the NMR spectra and relaxation rates, it is important to carefully control the orientation of $\mathbf{B}_0$ with respect to the unit cell of the crystal. The sample was secured with superglue inside the coil to prevent motion due to torque arising from the anisotropic susceptibility. The coil was attached to a platform that was in turn inserted in a dual-axis goniometer with rotation axes $\hat{\bm{\alpha}}$ and $\hat{\bm{\beta}}$ as illustrated in Fig. \ref{fig:align}(a).
Nominally, the $c$-axis should be parallel to $\hat{\bm{\alpha}}$ with $\mathbf{B}_0 \perp \hat{\bm{c}}$, such that $\beta$ controls the angle $\theta$ between $\mathbf{B}_0$ and the crystal $c$-axis, and $\alpha$ controls the angle $\phi$ between the projection of $\mathbf{B}_0$ in the plane and the $[110]_T$ direction of the orthorhombic distortion (see Fig. \ref{fig:align}(c)). If there is a misalignment angle, $\gamma$, between $\hat{\bm{c}}$ and $\hat{\bm{\alpha}}$, then the relationship between the angles $(\theta, \phi)$ and $(\alpha,\beta)$ becomes more complex. The condition $\mathbf{B}_0 \perp \hat{\bm{c}}$ is satisfied for:
\begin{eqnarray}\label{eqn:alphabeta}
  \sin\gamma \sin^2\alpha \cos(\alpha +\phi)&=&-\cos\alpha
   \left[\cos\gamma\sin\beta + \sin\gamma \right.\\
 \nonumber  && \left.(\cos\beta \cos\phi -\sin\alpha\sin(\alpha
   +\phi)\right].
\end{eqnarray} The misalignment can point in any direction around the axis determined by $\hat{\bm{\alpha}}$.
Rotation of the sample platform around the direction $\hat{\bm{\beta}}$ can restore $\mathbf{B}_0 \perp \hat{\bm{c}}$, however this $\beta$ correction changes every time the sample is rotated around $\hat{\bm{\alpha}}$ to adjust the in-plane angle $\phi$. If there were no misalignment ($\gamma=0$), then no $\beta$ corrections would be necessary.  Due to the strong anisotropy in the relaxation rate we can identify the in-plane orientation for any $\alpha$ with a precision $<0.1^{\circ}$ \cite{Wang2021}. Thus by rotating $\beta$ at each value of $\alpha$ to find the minimum value of \slrr, we can find the condition $\theta = 90^{\circ}$ and $\phi = \alpha$, as illustrated in Fig. \ref{fig:align}(b). We have verified that there is a one-to-one correspondence between changes in $\alpha$ and $\phi$, such that $\phi=\alpha +\alpha_0$, where $\alpha_0$ is a constant that is determined by the initial unknown in-plane orientation on the sample. By tracking the functional dependence $\beta(\alpha)$ we can reconstruct the initial misalignment angle, $\gamma$, which in principle allows us to rotate the sample into any orientation, provided that one accounts carefully for the backlash of the goniometer. We have further adjusted $\alpha$ such that $\phi=0$ corresponds to the $[110]_T$ direction corresponding to the $B_{2g}$ distortion in the ferroquadrupolar phase.

\subsection{Electric Field Gradient\label{sec:spectra}}

\begin{figure*}
\includegraphics[width = \linewidth]{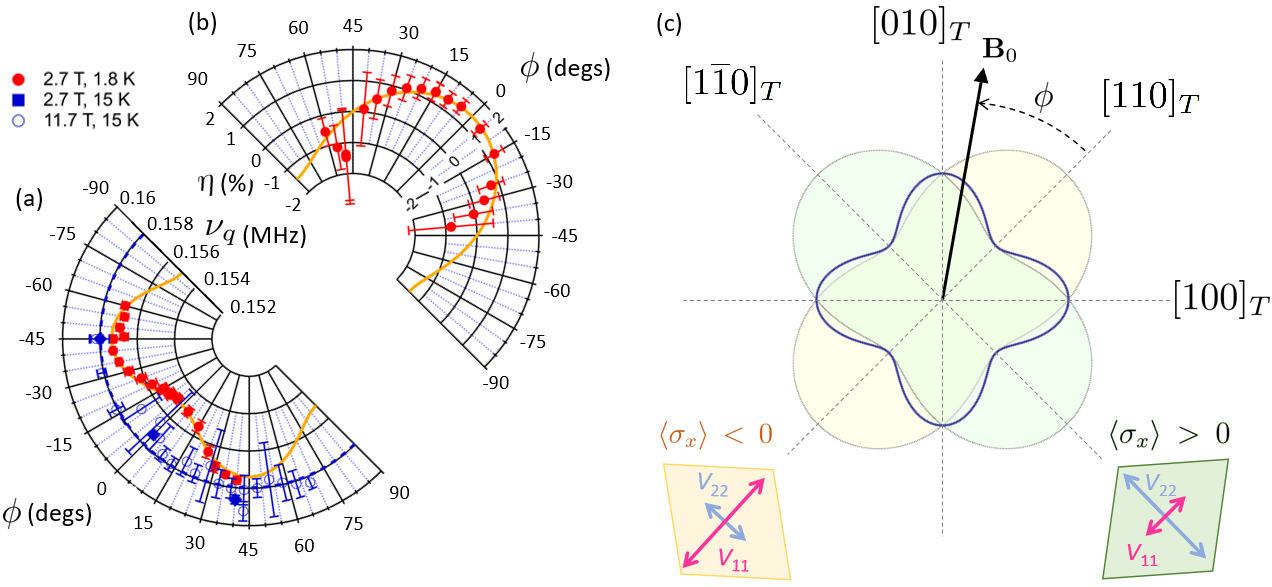}
\caption{(a) Polar plot showing the average quadrupolar splitting, $\nu_q$, as a function of in-plane angle, $\phi$, at 2.7 T and 1.8 K (solid red circles), 2.7 T and 15 K (solid blue squares) and 11.7 T and 15 K (open blue circles). The solid yellow line is a fit as described in the text, and the dashed line is a guide to the eye assuming tetragonal symmetry with no in-plane anisotropy ($\eta=0$).
(b) The $\phi$-dependence of $\eta$ at $B=2.7$~T and $T=1.8$~K. The solid yellow line is given by $\eta(\phi)=\eta_0\cos(2\phi)$ with {$\eta_0=1.44$\%}. Thus $\eta(\phi)$ switches sign from positive to negative as the field is rotated from $0^{\circ}$ to $90^{\circ}$. (c) Polar plot of $\nu_q$ as a function of field direction for constant $\eta > 0$ (green) and $\eta<0$ (yellow). The thick blue line shows $\nu_q(\phi)$ with angular-dependent $\eta(\phi)$.  The diamonds indicate the two domains with negative shear strain, $\varepsilon_{xy}$ (order parameter $\langle \sigma_x\rangle < 0$, yellow) and positive strain ($\langle \sigma_x\rangle > 0$, green). $\phi$ measures the angle relative to the $[110]_T$ direction in the tetragonal unit cell, such that $\phi=0$ lies along the $B_{2g}$ distortion direction. {Note that $\eta \propto \varepsilon_{xy} \propto  \langle \sigma_x\rangle $}.}
\label{fig:EFGsummary}
\end{figure*}

The left panel of Fig. \ref{fig:waterfall} displays spectra as a function of in-plane field direction below $T_Q$ at $\theta=90^{\circ}$ at 1.8 K and 2.7 T.  There are seven transitions at frequencies given by $\nu = \gamma B_0 (1 + K) + n\nu_q$, where $\gamma = 11.193$ MHz/T is the gyromagnetic ratio, $K$ is the magnetic shift ($\approx -0.4\%$), $n = -3, \cdots ,+3$, and $\nu_q$ is the quadrupolar shift. To first order in the nuclear quadrupolar interaction, $\nu_q$ is given by:
\begin{equation}\label{eqn:nuq}
  \nu_q = \frac{\nu_{zz}}{2}  \left[3\cos^2\theta - 1+ \eta  \sin ^2\theta \cos
   (2 \phi )\right],
\end{equation}
where $\nu_{zz} = eQV_{33}/{12 h}$,
$\eta = (V_{22} - V_{11})/(V_{11}+V_{22})$,
and $V_{ii}$ ($i = 1,2,3$) are the eigenvalues of the principal directions of the EFG tensor \cite{CPSbook}.  {In the disordered tetragonal phase, we find that {$\nu_q$} is independent of $\phi$, reflecting the axial symmetry of the EFG where $V_{11}= V_{22}$ and $\eta=0$. Below $T_Q$ there is a small but discernible variation in $\nu_q$ as a function of $\phi$, as seen in Fig. \ref{fig:waterfall}. In this case $V_{11} \neq V_{22}$, and the principal directions of the EFG tensor lie along $[001]_T$ with eigenvalue $V_{33}$, along $[110]_T$ with eigenvalue $V_{11}$, and along $[1\overline{1}0]_T$ with eigenvalue $V_{22}$. Note that these axes lie at 45$^{\circ}$ with respect the high temperature tetragonal axes, as illustrated in Fig. \ref{fig:EFGsummary}(c). Moreover, {$\eta \propto \varepsilon_{xy}$}, where $\varepsilon_{xy}$ is the $B_{2g}$ strain in the ordered state. Fig. \ref{fig:EFGsummary}(a) displays $\nu_q$ as a function of $\phi$ above (blue) and below (red) $T_Q$. The spectra are fitted with Gaussian peaks and $\nu_q$ is determined from the average separation of the satellite peaks. Note that the separation of adjacent peaks is not exactly $\nu_q$ due to second order quadrupole shifts, however, the separation of corresponding high ($n>0$) and low frequency satellites ($n<0$) are equal to $2n\nu_q$, as shown in Fig. \ref{fig:waterfall} for illustration, so the average separation with smallest error bars is $\nu_q \equiv (2\nu_q+4\nu_q+6\nu_q)/12$.} It can be seen that  $T$ dependence of $\nu_q$ is weak at a particular orientation (namely $\phi= 45^{\circ}$) for which {$\nu_q$} is nearly identical at $1.8$~K and $15$~K. This behavior arises because the temperature dependence of {$\nu_q$} comes primarily from $\eta (T)$, and $\cos(2\phi)=0$ for $\phi= 45^{\circ}$, allowing to identify the directions of the principal axes in the plane.

The angular dependence of $\nu_q$ at 1.8 K observed in Fig. \ref{fig:EFGsummary}(a) exhibits four-fold symmetry, in contrast to the expectation from Eq. \ref{eqn:nuq}. {This fact makes it impossible to distinguish between the $V_{11}$ and $V_{22}$ directions and consequently the angles $\phi=0^{\circ}$ and $90^{\circ}$ are assigned arbitrarily.} Moreover, the spectra in Fig. \ref{fig:waterfall} reveal only one set of peaks below 2 K, suggesting that there is a single nematic domain. These observations suggest the presence of a magnetoelastic coupling such that $\mathbf{B}_0$ detwins the ferroquadrupolar order causing $\eta$ to depend on the field direction. A similar phenomenon affects the nematicity in the iron based superconductors \cite{Chu2010fielddetwinning}.  If we assume $\eta = \eta_0\cos (2\phi)$, then we are able to model the data in Fig. \ref{fig:EFGsummary}(a) by Eq. \ref{eqn:nuq} with {$\eta_0=1.44$\%}, as  illustrated in Fig. \ref{fig:EFGsummary}(b,c). This behavior suggests the Tm quadrupole moments rotate with the magnetic field direction, such that domains with positive shear strain are always lower in energy. {The fact that $\eta(\phi)$ changes sign implies that the principal axes of the EFG switches, however we assign a fixed  axis as the evolution of $\nu_q(\phi)$ is captured well by a smoothly varying asymmetry parameter.} Although the non-Kramers doublet has $g_{ab} = 0$ and the Zeeman interaction vanishes to first order for in-plane fields, the doublet does couple quadratically to $\mathbf{B}_0$ as discussed below in section \ref{sec:SecondZeeman}. In effect, sufficiently large fields $\mathbf{B}_0$ induces Tm moments, which can orient the nematic domains.
At low fields and temperatures we are able to accurately fit the full spectra (see left panel of Fig. \ref{fig:waterfall}) for all orientations with two magnetic sites but equal EFG parameters $\nu_{zz}$ and $\eta$.  In principle, we might expect two sites with $\pm\eta$ corresponding to different nematic domains, at least for some angles. However,  even if there were two sites with different $\eta$ in the nematic phase in zero field, as discussed below the magnetoelastic coupling lowers the free energy of one domain over the other, effectively giving a single EFG environment for $\phi \neq 45^{\circ}$. Moreover, $\nu_q$ is independent of $\eta$ for $\phi=45^{\circ}$ (see Eq. \ref{eqn:nuq}), so even if two nematic domains were present their spectra would overlap at this angle. The equal width for all satellites indicates that there is little to no inhomogeneity of the EFG parameters. This result indicates both that any variations of $\eta$ in between nematic domains and the structural disorder due to any impurities, crystal defects and their associated strains, must be small.

\subsection{Second Order Zeeman Effect\label{sec:SecondZeeman}}
The low energy physics of \tmvo\ can be understood in terms of an Ising-type coupling between the Tm moments in the lattice \cite{Gehring1975},
\begin{equation}
    \mathcal{H}_Q = -\sum_{i\neq j} J_{ij}\hat{P}_{xy}(i)\hat{P}_{xy}(j),
    \label{eqn:exchange}
\end{equation}
where $\hat{P}_{xy} = (\hat{J}_x\hat{J}_y + \hat{J}_y\hat{J}_x)/2$ is the quadrupole operator for $B_{2g}$ strain at the $i^{th}$ site, and $\hat{J}_{\alpha}$ are the $J=6$ spin operators of the Tm \cite{FisherNematicQCP}.
Within the ground state doublet manifold this operator can be written as $\hat{P}_{xy} = E\hat{\sigma_{x}}$, where $\hat\sigma_{\alpha}$ are the Pauli matrices, and $E=21/2f^2 + 6\sqrt{10}fg + \sqrt{165}eg$
for the ground state doublet with wavefunction $|\psi_{1,2}\rangle = e|\pm 5\rangle + f|\pm 1\rangle + g|\mp 3\rangle$ in the $J_z$ basis \cite{FisherNematicQCP}\footnote{Note that reference \cite{FisherNematicQCP} considers a $B_{1g}$ distortion, however the physics is identical to that of \tmvo\ with a 45$^{\circ}$ rotation in the $xy$ plane.}.   The $J_{ij}$ are determined by the coupling between the lattice strain and the Tm quadrupolar moments, and gives rise to the ferroquadrupolar ordering at $T_Q = 2.15$ K in the absence of any external magnetic or strain fields. The EFG asymmetry, $\eta$, is proportional to the quadrupole order parameter $\langle \hat{P}_{xy}\rangle = E\langle \hat\sigma_x\rangle$. A magnetic field, $B_z$, oriented along the $c$-axis couples linearly to each Tm doublet as $\mathcal{H}_{Z,\parallel} = g_c\mu_B B_z\hat\sigma_z$. $B_z$ is thus a transverse field to the Ising order parameter, and suppresses the long range order at a quantum phase transition at $B_c = 0.515$ T \cite{FisherNematicQCP,Massat2021}.

\begin{figure}
\centering
 \includegraphics[width=\linewidth]{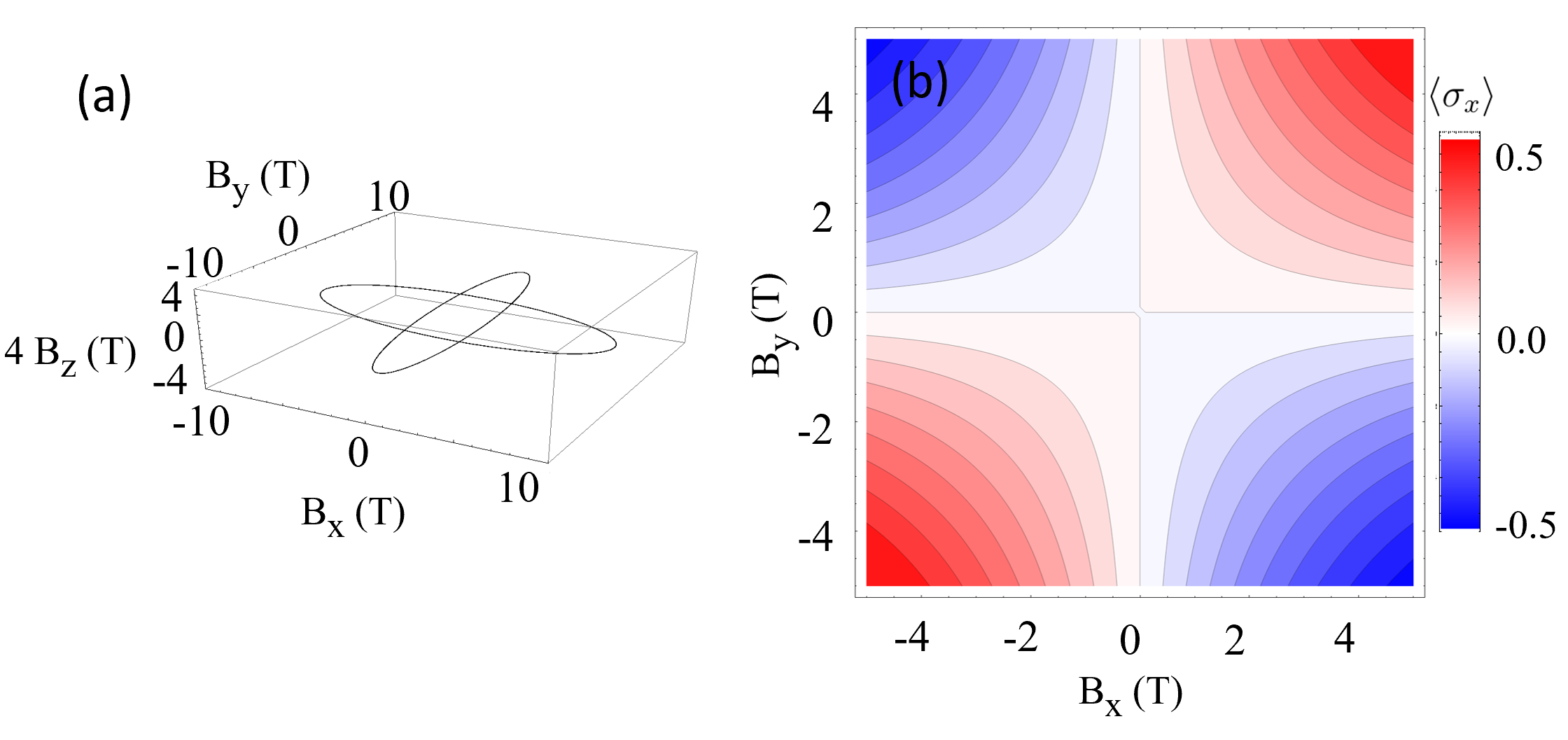}
  \caption{(a) Quantum critical field as a function of magnetic field orientation.  (b) Longitudinal fields induce finite order above $T_Q$ and smear out the phase transition.  The ferroquadrupolar order parameter, $\langle \sigma_x\rangle$, is shown at $T=4$ K $>T_Q$  as a function of field direction in the plane perpendicular to the $c$ axis. The order parameter vanishes only for $B_x = 0$ or $B_y=0$.}
  \label{fig:phase_diagram}
\end{figure}

For magnetic fields perpendicular to the $c$-axis, the $g$-factor is zero and the Zeeman interaction vanishes to first order \cite{Griffith1963}. However, for finite fields the admixture of excited crystal field levels gives rise to a second order Zeeman interaction \citep{Washimiya1970}:
\begin{equation}
\mathcal{H}_Z = -\frac{(g_J\mu_B B)^2}{2}[a\hat{\sigma}_1+b\cos(2\phi)\hat{\sigma}_x+c\sin(2\phi)\hat{\sigma}_y],
\label{eqn:zeeman}
\end{equation}
where $\hat{\sigma}_1$ is the identity matrix,  $g_J = {7}/{6}$ is the Land\'{e} factor, $B$ is the magnitude of the in-plane field, and $\phi$ is defined in Fig. \ref{fig:align}(c).  The parameters $a$, $b$, and $c$ can be determined by perturbation theory and depend on the crystal field levels of the Tm, as shown in \cite{BleaneyTmVO4review}. Note that for $\phi = 0$ or $90^{\circ}$ {($\mathbf{B}_0\parallel [110]_T$)} the magnetic field should act as a longitudinal field to the ferroquadrupolar order, wiping out the phase transition at $T_Q$.  For $\phi = 45^{\circ}$ {($\mathbf{B}_0\parallel [100]_T$)}, the magnetic field acts as a transverse field, such that $T_Q$ is suppressed, but the transition should remain sharp. Using the values $b=c = a/2 = 0.082$ K$^{-1}$ reported by Bleaney \textit{et al.}, we estimate that $T_Q$ should be suppressed to zero at a critical in-plane field of $B^*\approx \sqrt{2 T_Q/c}/g_J \mu_B \approx 9.2$ T along this direction.  Note that for any other in-plane angles, the phase transition will be smeared out.  Thus there exist two intersecting lines of quantum phase transitions as as function of field orientation, as illustrated in Fig. \ref{fig:phase_diagram}(a).

For $B=2.7$ T where the angular dependence of the EFG asymmetry parameter was measured (Fig. \ref{fig:EFGsummary}(b)), ${b}(g_J\mu_B B)^2/2\approx {c}(g_J\mu_B B)^2/2\sim 0.1 k_B T_Q$ thus the field should have a minimal effect on the ferroquadrupolar order.  On the other hand, the Zeeman interaction should lower the energy of one domain over the other, giving rise to a field-induced detwinning in the ordered state and the subsequent anisotropy in $\eta$ observed in Fig. \ref{fig:EFGsummary}(b). In other words, as the field is rotated from $\phi=0$ to $\phi=90^{\circ}$, domains with {positive} $\langle \sigma_x\rangle$  will convert to domains with {negative} $\langle \sigma_x\rangle$, in a manner similar to ramping an external magnetic field from negative to positive in a ferromagnet, and illustrated in Fig. \ref{fig:phase_diagram}(b).  An important difference, however, is that any field component  away from $\phi = \pm 45^{\circ}$ or $\pm 135^{\circ}$ will act as a longitudinal field, giving rise to a crossover rather than a sharp phase transition.  Moreover, these fields will favor one domain over the other, leading to a single value of $\eta$, as discussed above.

\subsection{Spectra in High Fields}

Spectra at the highest investigated field $B=11.7294$~T reveal the presence of two $^{51}$V sites with different magnetic shifts, as shown in the left panel of Fig. \ref{fig:waterfall}. The anisotropy in the magnetic shifts appears to have an even higher periodicity than $4\phi$ since spectra for $\phi \approx 45^{\circ}$ and $\phi \approx -45^{\circ}$ are not identical.  The origin of these two sites remains unclear.  There is only one V site per unit cell, even in the presence of a magnetic field. It may be that the magnetoelastic couplings may slightly perturb the principal axes of the magnetic shift tensor due to distortions of the VO$_4$ tetrahedra.

The existence of two magnetic sites has been remarked in earlier studies but could not be explained by the formation of two nematic domains since the area ratio of the two associated peaks has an unexpected dependence on field orientation \cite{BleaneyTmVO4review}. For $\phi \neq 45^{\circ}$ and at low fields the two magnetic sites become apparent only below $T_Q$, but in the highest applied field of $B \approx 11.7$~T the splitting in the magnetic shift survives far beyond 15~K. This behaviour could be expected for a signature of nematicity that survives due to a large longitudinal field.

\section{Spin Lattice Relaxation}

\subsection{Multiple Relaxation Channels\label{sec:relax}}

An important consequence of Eq. \ref{eqn:zeeman} is that in-plane fields induce Tm magnetic moments (linear in $B$) which couple to the field itself. These induced moments can couple to the nuclei through the hyperfine interaction \cite{Wang2021,Washimiya1970}.  $^{51}$V has both a nuclear magnetic moment and a nuclear quadrupolar moment ($Q=52$ millibarn), and therefore  both fluctuations of the hyperfine field and fluctuations of the EFG can give rise to spin lattice relaxation. In a previous study we investigated the anisotropy of the nuclear spin lattice relaxation rate and found evidence that fluctuations of the Tm quadrupoles may contribute to the nuclear spin lattice relaxation via the EFG at the V site \cite{Wang2021}. In general, there are three distinct relaxation channels: a magnetic, $W_m$, and two quadrupolar relaxation rates, $W_{Q1}$ and $W_{Q2}$, but it is difficult to disentangle the contribution of each \cite{suterquadrupolarrelaxation}. These relaxation channels couple different sets of the $I_z$ nuclear spin levels. The relaxation measured at a particular nuclear spin transition $m\leftrightarrow (m-1)$ is a complicated function of $W_m$, $W_{Q1}$ and $W_{Q2}$ determined by a master equation (with an $8\times 8$ dimensional matrix for spin $7/2$ of $^{51}$V), as discussed in Appendix \ref{sec:AppendixMas}. The relaxation function for each transition is slightly different, thus by measuring the relaxation at multiple transitions one can globally fit the set of seven recovery curves to extract $W_m$, $W_{Q1}$ and $W_{Q2}$.   However, in order to do so it is vital to resolve and excite each transition in the spectrum individually. The inhomogeneous broadening from the demagnetization field precluded such studies previously, but fortunately all seven transitions are clearly visible in the FIB sample.

\begin{figure}
\includegraphics[width=\linewidth]{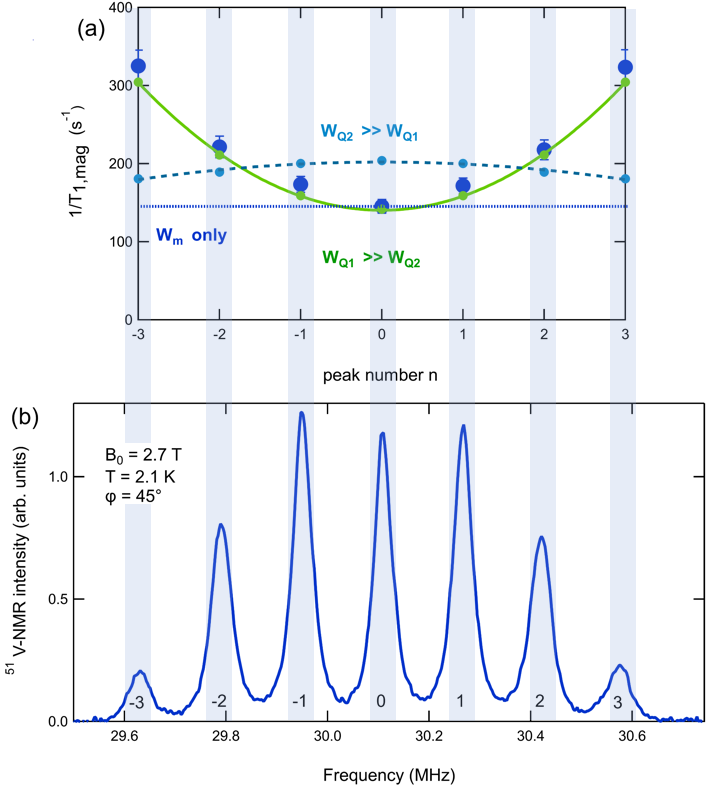}
\caption{\label{fig:relaxation_channels} (a) Plot of the magnetic relaxation rate $T^{-1}_{1,\rm{mag}}=2W_m$ (large blue circles) with strongly enhanced relaxation rates on the outermost satellites ($n=\pm3$) at $B_0=2.7$~T and $T=2.1$~K. Purely magnetic relaxation should give identical $T^{-1}_{1,\rm{mag}}$ for all peaks (horizontal line). Fits to magnetic relaxation only of simulated relaxation curves that include $W_{Q1}=2.3$~s$^{-1}$ and $W_{Q2}=0$ reproduce the upward curvature of $T^{-1}_{1,mag}$. Inclusion of $W_{Q2}=2.3$~s$^{-1}$, but $W_{Q1}=0$ leads to downward curvature. Lines are quadratic fits and vertical bars link relaxation data with the corresponding transitions of the NMR spectrum. (b) Spectrum of FIB sample at $B=2.7$~T and $T=2.1$~K. }
\end{figure}

Fig. \ref{fig:relaxation_channels} shows values obtained at each of the seven peaks by fitting the measured relaxation curves by the formulae corresponding to the magnetic relaxation only.
If there were only magnetic fluctuations, all seven peaks would be described by a single value of $T_{1,mag}^{-1} = 2W_m$, shown by the dotted line in Fig. \ref{fig:relaxation_channels}(a). Such a model clearly does not fit the data, which reveal a significant enhancement of the relaxation of the satellites. On the other hand, including a finite $W_{Q1}$ accurately captures the trend visible in the data, illustrated by the solid green line in Fig. \ref{fig:relaxation_channels}(a). Here, we have computed the relaxation curve for each transition in the presence of both $W_m$ and $W_{Q1}$, and fit the curves to a purely magnetic relaxation model. Thus, the enhanced relaxation at the satellites indicates the presence of EFG fluctuations that contribute to the nuclear spin relaxation. Including $W_{Q2}$ has a minor effect that is opposite to what is observed: if $W_{Q2}$ fluctuations dominate, then the relaxation is enhanced for the inner transitions and suppressed for the outer transitions, as illustrated by the dashed cyan line. Thus our data suggest that $W_{Q1}>W_{Q2}$. This behavior is expected for dominantly $B_{2g}$ fluctuations, as $W_{Q1}(\theta)$ and $W_{Q2}(\theta)$ are transformed when the field is applied perpendicular to the principal EFG axis (along $[001]$), where we expect $W_{Q1}/W_{Q2}(\theta=90^{\circ})=4$, where $\theta$ is the angle between $\mathbf{B}_0$ and the $c$ axis \cite{Wang2021}. We use this relation as a constraint and perform global fits based on all seven relaxation curves including both magnetic and quadrupolar relaxation contributions.

Fig. \ref{fig:heatmap_45deg} shows the field and temperature dependence of $W_{Q1}/T$ and $W_{m}/T$ obtained by global fits to all seven nuclear transitions for $\phi=45^{\circ}$. Note that the magnetic relaxation rate ($W_m$) is between one to two orders of magnitude larger than the quadrupolar relaxation rate ($W_{Q1}$), which is partly a simple consequence of numerical values in the corresponding matrices (A2)-(A4) given in the Appendix A, where magnetic terms are one order of magnitude smaller than the quadrupolar ones. Information from Fig. \ref{fig:relaxation_channels} and Fig. \ref{fig:heatmap_45deg} allows us to compare the relative strength of these rates at $B=2.7$~T and $T=2.1$~K for $\phi=45^{\circ}$. Clearly, the enhancement of $T^{-1}_{1,\rm{mag}}$ on all satellite peaks must be due to substantial $W_{Q1}$, but on the central peak we found $T^{-1}_{1,\rm{mag}}=145.5$~s$^{-1}$ and based on the global fits this value is close to $2W_m=138.0$~s$^{-1}$. Thus $W_{Q1}$ is responsible for only a minor enhancement of $\sim 5$~\% on the central peak, which is consistent with Ref.~\citep{suterquadrupolarrelaxation}. Nevertheless, $W_{Q1}$ is not negligible as it doubles the relaxation rate on the outermost satellites. This observation implies that information about quadrupolar relaxation cannot be extracted from the central peak relaxation alone.

\subsection{Transverse Fields}
\begin{figure*}
  \includegraphics[width=\linewidth]{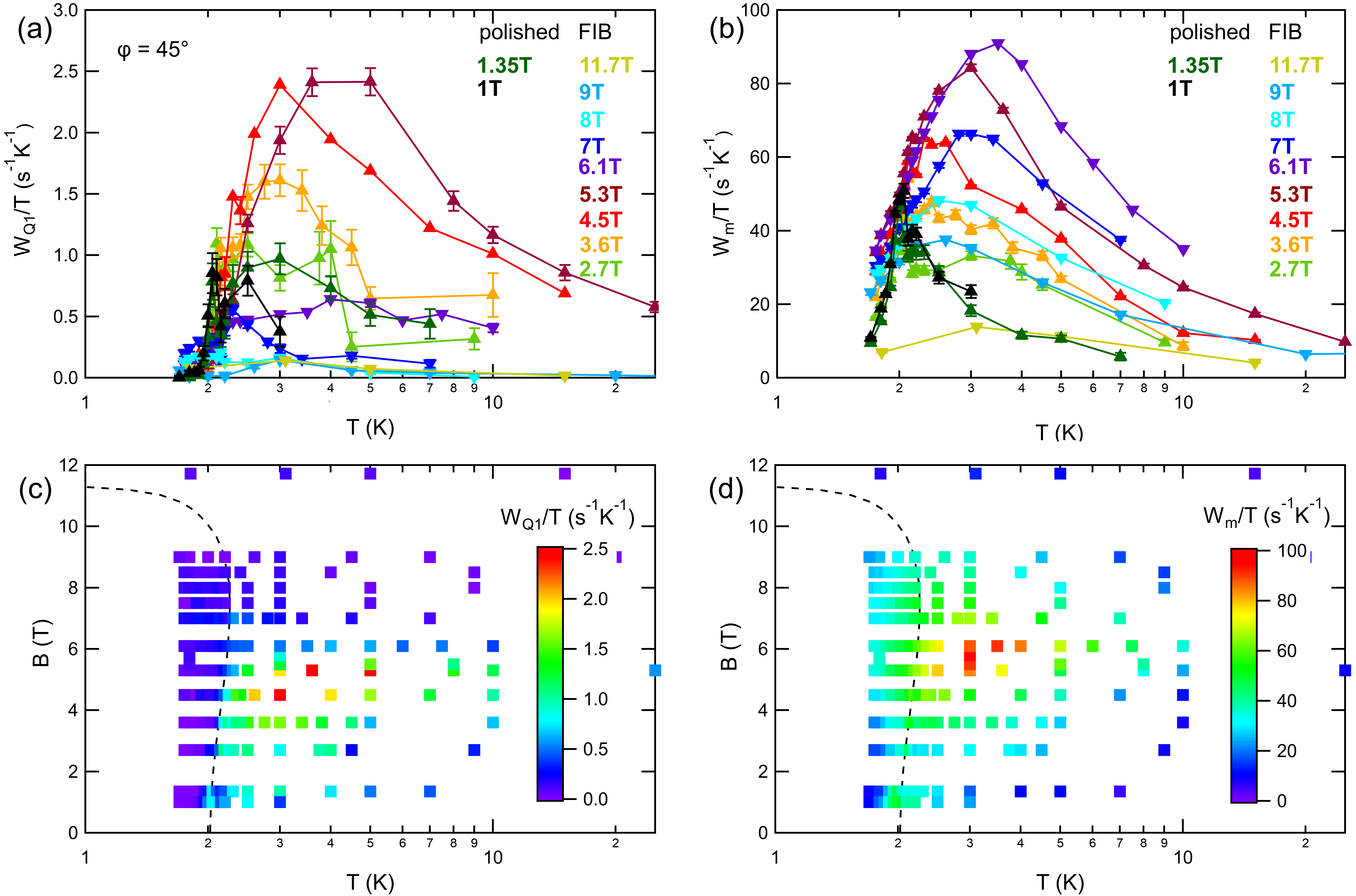}
  \caption{\label{fig:heatmap_45deg}(a) and (b): The quadrupolar and magnetic relaxation rates divided by temperature, $W_{Q1}/T$, $W_{m}/T$ at fields between 1~T and 11.7~T at $\phi=45^{\circ}$. $B<2.7$~T was measured on a different TmVO$_4$ sample that was polished by hand to a rounded shape. Below, in  panels (c) and (d), the same data is plotted as heat maps. At low fields, sharp anomalies at $T_Q=2.15$~K are visible and broaden and increase with increasing fields. Above $B \sim 6$~T both relaxation channels, especially $W_{Q1}/T$, are rapidly suppressed. The dashed lines in the background of panels (c) and (d) are guides to the eye based on the computed $T_Q(B)$ shown in Fig. \ref{fig:effectivephasediagram}. }
\end{figure*}

The data shown in Fig. \ref{fig:heatmap_45deg} were acquired for the transverse field orientation, in which $\mathbf{B}_0$ is aligned orthogonal to the ferroquadrupolar distortion (see Fig. \ref{fig:EFGsummary}(c)). Note that $W_{Q1}/T$ and $W_{m}/T$ are determined by the dynamical magnetic and nematic susceptibilities of the Tm 4f electrons:
\begin{eqnarray}
    \frac{W_m}{T} &\sim&
    \lim_{\omega \rightarrow 0} \sum\limits_{\mathbf{q},\alpha,\beta} \mathcal{F}_{\alpha\beta}(\mathbf{q}) \frac{\textrm{Im}\chi_{\alpha\beta}^{mag}(\mathbf{q},\omega)}{\omega} \\
     \frac{W_{Q1}}{T} &\sim& \sum\limits_{\mathbf{q}}  \frac{\textrm{Im}\chi^{nem}(\mathbf{q},\omega)}{\omega},
\end{eqnarray}
where $\mathcal{F}_{\alpha\beta}(\mathbf{q})$ is the hyperfine form factor \cite{DioguardiPdoped2015,Wang2021}. The dynamical magnetic and nematic susceptibilities, $\chi^{mag}$ and $\chi^{nem}$, are likely coupled to one another and are driven by the fluctuations of the Tm quadrupolar moments. The details of the frequency dependence of the susceptibility and relaxation of the collective excitations, however, remain unclear.  Qualitatively, we expect contributions to arise both from the dynamics of isolated doublets split by the Zeeman interaction and coupled to a bath of phonons \cite{Leggett1987}, as well as the collective dynamics associated with the ferroquadrupolar ordering in the vicinity of $T_Q$ \cite{MelcherCJTEreview}.

\begin{figure}
  \includegraphics[width=\linewidth]{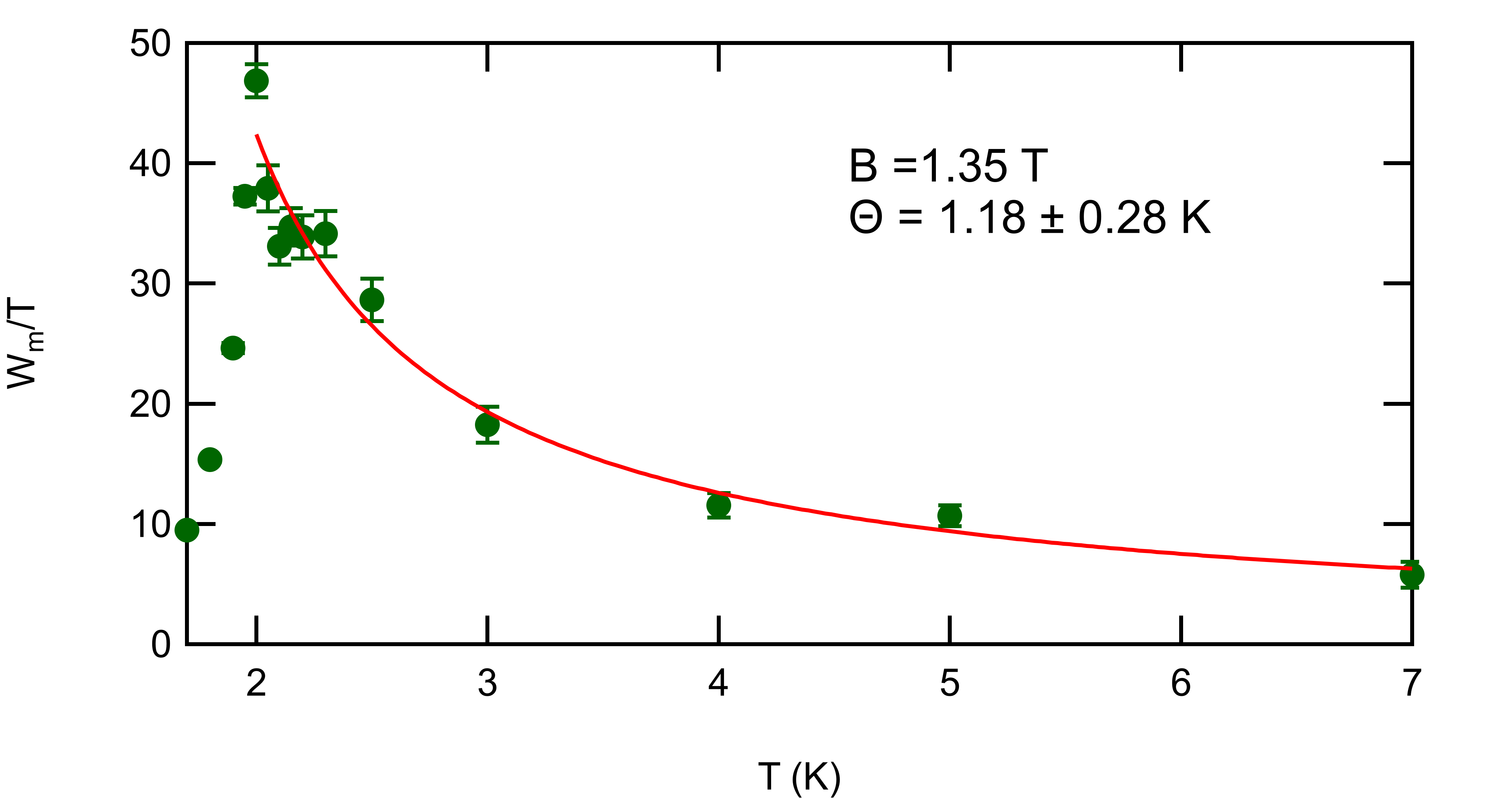}
  \caption{\label{fig:CWfit} The magnetic relaxation rate, $W_m/T$ versus $T$ for $\phi=45^{\circ}$ and $B_0$ = 1.35 T.  The solid red line is a fit to Eq. \ref{eqn:CW} as described in the text.}
\end{figure}

For the lowest fields (up to 1.35 T), both $W_{Q1}/T$ and $W_{m}/T$ exhibit a relatively sharp peak at $T_Q$ and are suppressed below in the ordered state. The relaxation rate above $T_Q$ can be fit to a Curie-Weiss expression:
\begin{equation}
    \frac{W_{m}}{T} = \frac{C_0}{T-\Theta}.
    \label{eqn:CW}
\end{equation}
with a Weiss temperature $\Theta/T_Q =0.55\pm 0.13$, as shown in Fig. \ref{fig:CWfit}. This behavior likely reflects the divergence of nematic susceptibility at the phase transition \cite{FisherScienceNematic2012,DioguardiPdoped2015}. The fact that $\Theta/T_Q < 1$ indicates that the elastic coupling between the Tm 4f moments and the $\varepsilon_{xy}$ lattice strain renormalizes the exchange interaction between the moments \cite{FisherScienceNematic2012}. Although we fit the data to only a limited temperature range limiting the precision of the fit, the value we obtain is consistent with previous measurements of the lattice elastic constant, $c_{66}$ \cite{TmVO4CJTE}.

\begin{figure}
  \includegraphics[width = \linewidth]{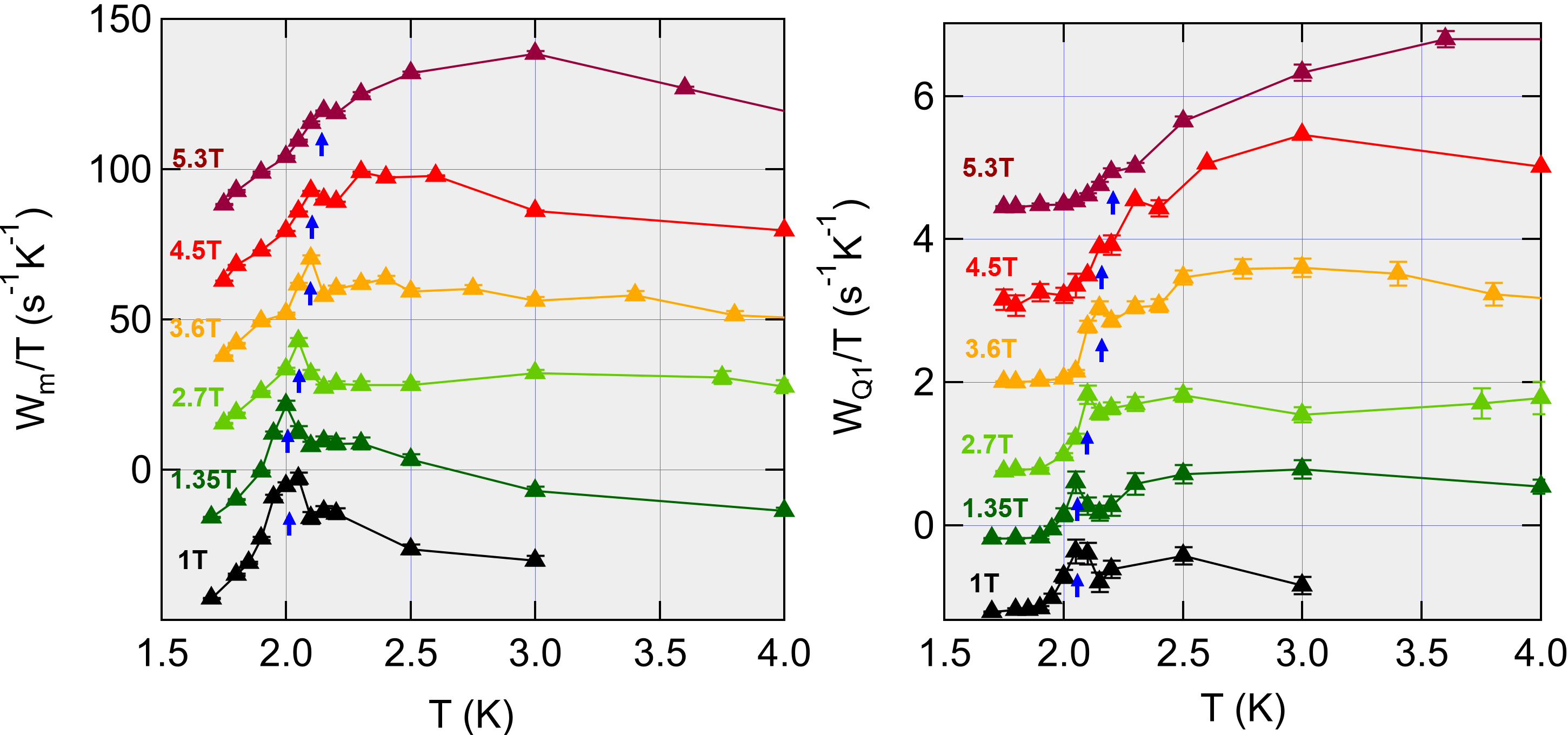}
  \caption{\label{fig:fieldenhancement} $W_m/T$ and $W_{Q1}/T$ versus $T$ for small fields along $\phi = 45^{\circ}$.  The data at each field have been displaced vertically for clarity. Blue arrows are guides to the eye indicating $T_Q$.}
\end{figure}

As the magnitude of the field increases, the sharp feature shifts upwards in temperature, and a large broad peak emerges around $\sim 3$~K, as shown by the blue arrows in Fig. \ref{fig:fieldenhancement}. The peak temperature increases by approximately 0.2 K by 5.3 T, and then disappears below the broad peak. The behavior up to this field suggests that the transition temperature $T_Q$ increases with field, in contrast to the expectation that transverse fields should suppress $T_Q$ to zero monotonically. The enhancement we observe may reflect the influence of higher order crystal field levels, as discussed below in section \ref{sec:discuss}. At higher fields, both $W_{Q1}/T$ and $W_{m}/T$ exhibit broad maxima around 3~K, and the feature associated with $T_Q$ disappears. The origin of this behavior is unknown, but may be related to a change in the dynamics of isolated two-level systems as the Zeeman interaction increases. It is unclear if the ferroquadrupolar ordering is suppressed, or the 3~K feature simply overwhelms the signature of the phase transition.  For $B_0\gtrsim 6$ T, both relaxation rates decrease with increasing field, and reach their low-field values by approximately 9 T.  The peak at 6 T may reflect the critical in-plane field, $B^*$, which is close to the estimated value based on the CEF parameters as discussed above.  For higher magnetic fields, the splitting of the doublets will exceed $k_B T$ and the dynamics will be gapped out.

\subsection{Longitudinal fields}

\begin{figure}
 \includegraphics[width = \linewidth]{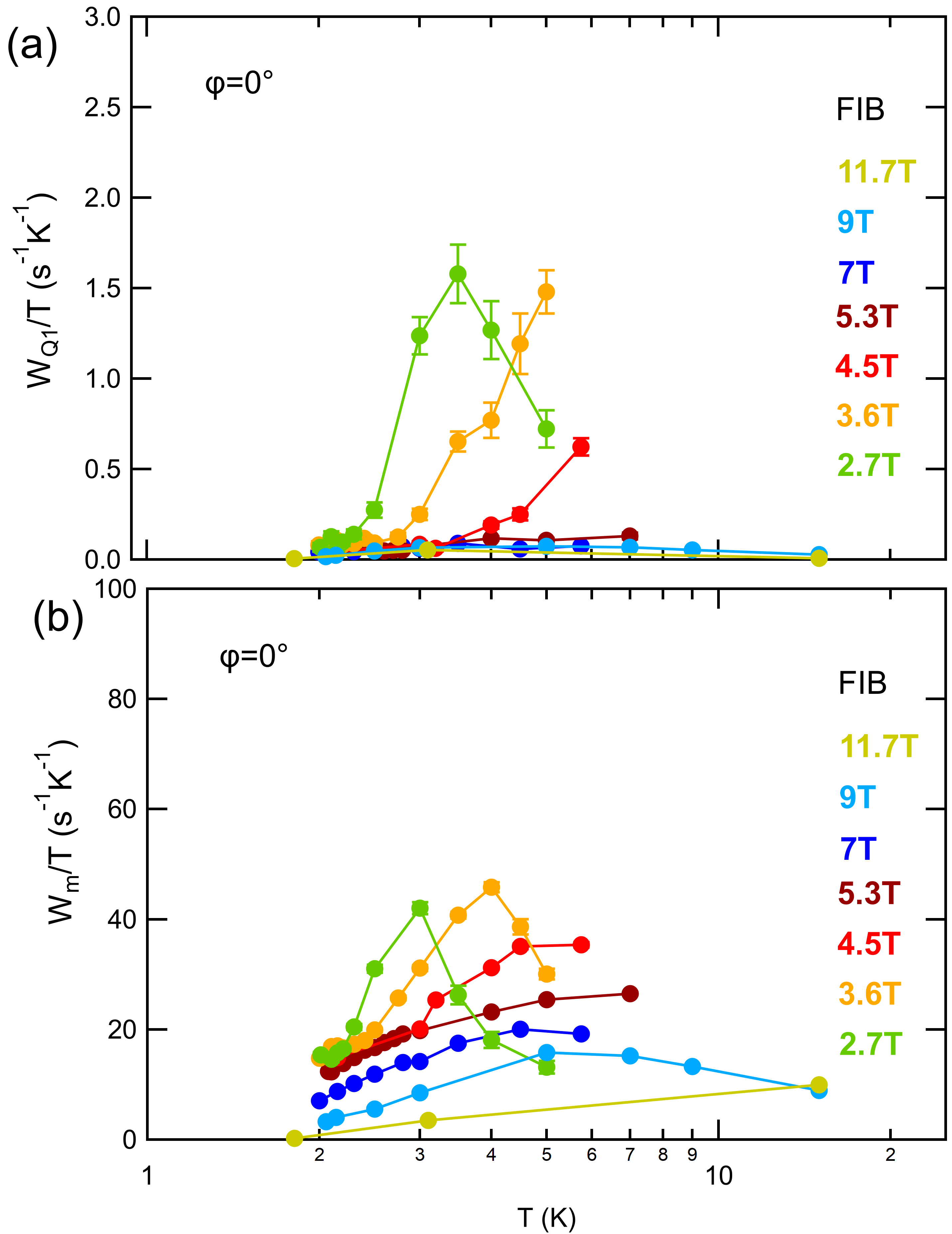}
  \caption{\label{fig:heatmap_0deg}(a) and  (b): Plots of the quadrupolar, $W_{Q1}/T$, and magnetic, $W_{m}/T$, relaxation rates divided by temperature for fields between 2.7~T and 11.7~T for $\phi=0^{\circ}$.
  At low fields, unlike for $\phi=45^{\circ}$, no sharp anomalies at $T_Q=2.15$~K are visible and the relaxation rates are always smaller.}
\end{figure}

Both $W_{Q1}/T$ and $W_{m}/T$ display a markedly different behavior for the longitudinal field orientation ($\phi=0^{\circ}$), as seen in Fig. \ref{fig:heatmap_0deg}.  There is no sharp peak at $T_Q$ in the investigated field range, as expected when longitudinal fields broaden the phase transition. There is, however, a broad peak with a peak temperature that increases strongly with field. Overall, the relaxation rates for both channels are smaller than in the transverse field direction. The fact that the broad peak around 3-4 K emerges for both transverse and longitudinal fields suggests that it is unrelated to the collective phase transition, but rather associated with isolated field-split doublets.

\subsection{Angular variation}

\begin{figure*}
\includegraphics[width=\linewidth]{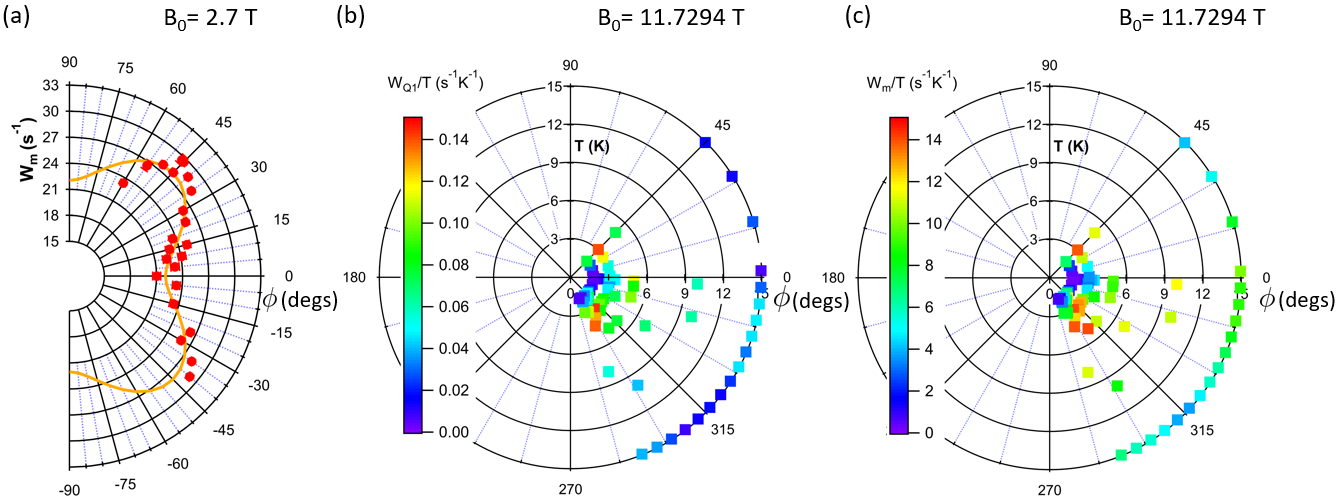}
    \caption{(a) The magnetic relaxation rate divided by temperature $W_m/T$ as a function of $\phi$ measured at $T=1.8$~K and $B_0=2.7$~T. Polar plots of quadrupolar fluctuations $W_{Q1}/T$ (b) and magnetic fluctuations $W_{m}/T$ (c) versus $\phi$ at $B_0$=11.7294~T, where the radius corresponds to the temperature. At lower temperatures $T<7$~K nematic fluctuations are strongest around $\phi=45^{\circ}$, but for $T>7$~K the anisotropy is reversed, possibly due to direct coupling to the lattice. }
    \label{fig:angular}
\end{figure*}

Figs. \ref{fig:heatmap_45deg} and \ref{fig:heatmap_0deg} reveal that at low temperature the quadrupolar fluctuations vanish for both orientations such that $W_{Q1}$ is negligible and the relaxation is determined entirely by $W_m$. Fig. \ref{fig:angular}(a) displays the angular dependence of $W_m$ at $T=1.8$~K and $B_0=2.7$~T.  This quantity exhibits a maximum at $\phi=45^{\circ}$, which likely reflects the fact that transverse fields enhance the ferroquadrupolar fluctuations. At 11.7 T, both the quadrupolar and magnetic relaxation rates maintain the $4\phi$ periodicity at all investigated temperatures at this field, as seen in Figs.\ref{fig:angular}(b,c).  However, it should be noted that the relaxation rates are no longer maximal around $\phi=45^{\circ}$ at 15~K, but instead peak around $0^{\circ}$.  This unexpected reversal of the anisotropy suggests that ferroquadrupolar fluctuations are suppressed around 7~K at this field and another field-induced anisotropy develops, possibly related to the intrinsic anisotropy of isolated doublets.

\section{Discussion\label{sec:discuss}}

\subsection{Field Dependence}

\begin{figure}
\centering
 \includegraphics[width=\linewidth]{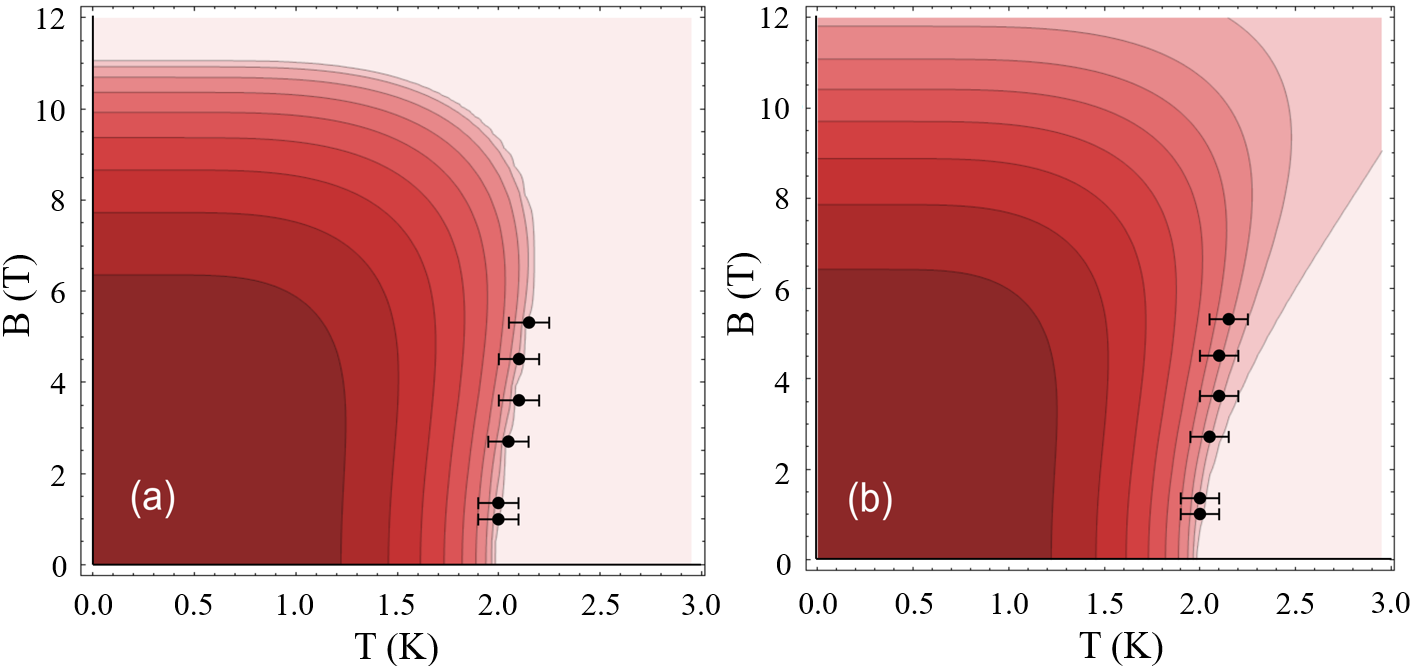}
  \caption{ Calculated phase diagram for the effective model given by Eq.~\ref{eqn:effective} with $\xi = 0.002$ $T^{-2}$. Panel (a) shows $\langle{\sigma}_x\rangle$ for $\phi=45^{\circ}$, and (b) shows the case for $\phi=44^{\circ}$.  Light colors correspond to $\langle{\sigma}_x\rangle = 0$ and darker red corresponds to increasing values of $\langle{\sigma}_x\rangle$. Solid black circles correspond to the peaks indicated by the blue arrows in Fig. \ref{fig:fieldenhancement}.}
  \label{fig:effectivephasediagram}
\end{figure}

The enhancement of $T_Q$ in transverse fields is surprising. A similar effect was observed in TmCd, a material that also experiences a cooperative Jahn-Teller effect with a phase transition at 3.16~K \cite{Luethi1973}.  However, the phase transition in TmCd is first order, and the effect was explained as the influence of a longitudinal field that shifts the first order discontinuity in the order parameter to higher temperatures and eventually broadens into a crossover. The phase transition in \tmvo\ is  second order, and the enhancement we observe occurs for transverse fields where the phase transition remains sharp. A possible explanation for this observation is that the quadrupole moment of the Tm ground state doublet is enhanced by the magnetic field.  As shown in Appendix \ref{sec:AppendixD}, the excited crystal field states admix with the ground state doublet, such that for small fields $E(B) = E(0) + \xi B^2$.  As a result, the behavior of the Tm moments described by Eqs. \ref{eqn:exchange} and \ref{eqn:zeeman} for transverse fields ($\phi = 45^{\circ}$) becomes:
\begin{equation}
    \mathcal{H}_{eff}(B) = -\sum_{i\neq j} J_{ij}E^2(B)\hat{\sigma}_{x,i}\hat{\sigma}_{x,j}  -c \frac{(g_J\mu_B B)^2}{2}\hat{\sigma}_y
    \label{eqn:effective}
\end{equation}
so $B$ both enhances the effective coupling between the Tm moments as well as induces quantum fluctuations. The mean field phase diagram for this model is displayed in Fig. \ref{fig:effectivephasediagram}(a) for $\xi=0.002$ T$^{-2}$.  As $B$ increases, $T_Q$ initially rises by $\sim 0.2$ K around 8 T and is then suppressed to zero at a critical field $\sim 11$ T.  The phase transition remains sharp for all fields. This model captures the enhancement of $T_Q$ for small $B$ observed in Fig. \ref{fig:fieldenhancement}.

The behavior at higher fields remains unexplained. The data in Fig. \ref{fig:heatmap_45deg} reveal a broad peak emerging around 3 K for $B\gtrsim 4$ T that spans several K, and the sharp feature associated with $T_Q$ at lower $B$ disappears, so that it is not possible to track the phase transition to higher fields. It is possible that the relaxation dynamics are simply dominated by the behavior of isolated field-split doublets, such that the collective behavior of the Tm moments is not evident at these high fields.

An alternative explanation is that there is a small misalignment from $\phi=45^{\circ}$.  As seen in Eq. \ref{eqn:zeeman}, such a misalignment would contribute to a longitudinal field that will smear out the transition. This effect would increase with field, such that the phase transition remains relatively sharp at low fields, but is broadened at higher fields.  This behavior is illustrated in  Fig. \ref{fig:effectivephasediagram}(b) for a misalignment of $1^{\circ}$, which is about the precision of our goniometer. In this case, the phase transition remains sharp for small $B$, with $T_Q$ rising with increasing $B$.  At higher fields, the phase transition is smeared out, and the crossover behavior extends to higher temperature.  Note, however, that for fields in the range 2 T $\lesssim B \lesssim 5$ T, $W_m/T$ and $W_{Q1}/T$ exhibit both a broad peak as well as a sharper feature at $T_Q$, suggesting two separate phenomena.  The phase diagram for a misaligned field would not have two features, but rather a single phase transition that broadens and moves to higher temperatures with increasing field, as illustrated in Fig. \ref{fig:effectivephasediagram}(b). Thus a misalignment is not fully consistent with our observations.

It is interesting to note that both the quadrupolar fluctuations ($W_{Q1}/T$) and the magnetic fluctuations ($W_{m}/T$) exhibit qualitatively similar trends with temperature and field in Fig. \ref{fig:heatmap_45deg}. This similarity suggests that the field-induced Tm moments are coupled with the Tm quadrupole moments, so that the critical fluctuations of the latter can drive fluctuations of the former. There are important differences between the two relaxation channels, however. For example, it is clear that $W_{Q1}/T$ is more rapidly suppressed below $T_Q$ than $W_{m}/T$. This observation suggests that $W_{Q1}/T$ is a more direct measure of the freezing nematic fluctuations when the order parameter $\langle \sigma_x\rangle$ steeply grows below $T_Q$.

The exact mechanism that leads to the broad peak above $T_Q$ at high fields remains unclear.  Previous measurements indicate that the spin lattice relaxation rate is correlated with $c_{66}$ up to 100 K at 11.7 T, suggesting that nematic fluctuations dominate the relaxation of the V nuclei \cite{Wang2021}.  Our results are consistent with the previous measurements, but we have now disentangled the two relaxation channels.  Both the previous and current measurements reveal the presence of the broad peak.  A qualitative understanding of this peak may be that for temperatures $T > \Delta/k_B$, where $\Delta$ is the splitting of the Tm doublet, the relaxation rate is determined by the growth of the nematic susceptibility as the lattice softens. On the other hand, for $T < \Delta/k_B$, the fluctuations will be suppressed with decreasing temperature \cite{Leggett1987}.  These two effects may conspire to give rise to the broad maximum when $T\sim \Delta/k_B$, which itself is field-dependent. Comparison of NMR relaxation rates with other low-frequency dynamic probes, such as   ultrasound attenuation,  can be fruitful \cite{Frachet2021}, however there is no report of ultrasound experiments with fields applied in  the plane to allow direct comparison.

\subsection{Disentangling Relaxation Channels}

Our study proposes a conceptually easy mechanism to identify and disentangle magnetic and quadrupolar relaxation rates. Since the seminal work of Suter \textit{et al.} it has been generally assumed that standard relaxation rate measurements preclude accurately determinations of $W_m$, $W_{Q1}$ and $W_{Q2}$ \cite{suterquadrupolarrelaxation}, except in cases where $\nu_Q$ is sufficiently small that double-resonance experiments (SEDOR) can be performed \cite{Suter2000,Suter2000PRL}. However, as we show in \tmvo, a deviation in the magnetic relaxation rates on two transitions, e. g. the central peak and an outermost satellite (see Fig. \ref{fig:relaxation_channels}), is a direct indication that quadrupolar mechanisms are at play. This could be complicated if satellites are broad in frequency and possess inhomogeneous relaxation rates across the peak, but valuable information about charge ordering could be extracted from the median relaxation rates extracted from stretched magnetic fits for different transitions in inhomogeneous systems such as cuprates like La$_{2-x}$Sr$_{x}$CuO$_{4}$ \cite{Mitrovic2008,Arsenault2020}. This method is not directly applicable to the well-studied case of the nematic order in FeSe as both NMR active isotopes $^{57}$Fe and $^{77}$Se have no satellites since $I=1/2$. However, FeSe$_{1-x}$S$_x$ could be an interesting material with $^{33}$S having $I=3/2$. Since the natural abundance is small such samples would need to be enriched with $^{33}$S. It is expected that the relaxation rate of $^{33}$S, $^{33}W_{Q1,2}/T$ will display a very different behaviour than $(^{77}T_1T)^{-1}$, because in the former $^{33}$S is sensitive to the divergent nematic susceptibility as the structural transition $T_s$ is approached. Measurements of $W_{Q1}$ and $W_{Q2}$ above and below the superconducting $T_c$ are a direct measure of interplay between superconductivity and nematicity as well as the importance of nematic fluctuations for the superconducting pairing. Related experiments have been performed in ${\mathrm{BaFe}}_{2}({\mathrm{As}}_{1\ensuremath{-}x}{\mathrm{P}}_{x}{)}_{2}$ by comparing relaxation rates of $^{75}$As ($I=3/2$) and $^{31}$P ($I=1/2$) \cite{Dioguardi2016}, but the method described here allows a direct determination of $W_{Q2}$.

\section{Conclusion}

\tmvo\ is a model system for investigating nematic quantum criticality in an insulator, in which ferroquadrupolar ordering of Tm 4f orbitals is well-described by the transverse field Ising model for magnetic fields oriented along the crystalline $c$-axis \cite{Massat2021}. The response to in-plane fields has not been previously investigated, because the Zeeman interaction for the Tm non-Kramers doublets vanishes to first order in this configuration.  We have studied the ferroquadrupolar ordering for in-plane fields using NMR of a single crystal shaped by a FIB to eliminate inhomogeneous broadening effects, and uncovered pronounced in-plane anisotropy in the EFG and spin lattice relaxation rates. By measuring the relaxation curves at multiple nuclear transitions of the $^{51}$V nucleus, we disentangle magnetic and quadrupolar relaxation channels. We find that the phase diagram for in-plane fields is dominated by the second order Zeeman interaction, and that {the in-plane magnetic  field can act as an effective transverse or  longitudinal  field to the Ising nematic order, depending on direction, and leads to a marked in-plane anisotropy in both relaxation channels. Also, we find that even for the case where the in-plane field corresponds to an effective transverse field, nevertheless small values of the magnetic field initially enhance the ferroquadrupolar ordering temperature before eventually suppressing the long-range order. This effect is tentatively ascribed to competing effects of field-induced mixing of higher energy crystal field eigenstates and the destabilizing effects of field-induced quantum fluctuations. } For higher fields, the spin lattice relaxation is dominated by a broad peak in temperature and field, which may be due to the dynamics of isolated doublets split by magnetic field. The second order Zeeman interaction also gives rise to a magnetoelastic coupling, so that as the magnetic field rotates in the plane, the nematic domains switch. Our results establish in-plane magnetic fields as a novel approach to tune \tmvo\ to a quantum phase transition, and also establishes an important new method to disentangle magnetic and quadrupolar relaxation channels for high spin nuclei in a variety of condensed matter systems. Our results also indicate that \tmvo\ should exhibit rich nonlinear magnetization effects for in-plane fields.\\

\begin{acknowledgements}
We acknowledge helpful discussions with R. Fernandes, A. P. Mackenzie, T. Briol, and B. Ramshaw and  V. Taufour for susceptibility measurements. Work at UC Davis was supported by the NSF under Grants No. DMR-1807889 and PHY-1852581, as well as the UC Davis Seed Grant program. Crystal growth performed at Stanford University was supported by the Air Force Office of Scientific Research under award number FA9550-20-1-0252. P. M. was partially supported by the Gordon and Betty Moore Foundation Emergent Phenomena in Quantum Systems Initiative through Grant GBMF9068. M.D.B. acknowledges partial support from the Swiss National Science Foundation under project number P2SKP2 184069, as well as from the Stanford Geballe Laboratory for Advanced Materials (GLAM) Postdoctoral Fellowship program.
\end{acknowledgements}

\bibliography{TmVO4NMRbibliography}

\appendix
\section{Master Equation \label{sec:AppendixMas}}
\begin{figure}
\centering
\hspace*{-0.3cm}
  \includegraphics[width=\linewidth]{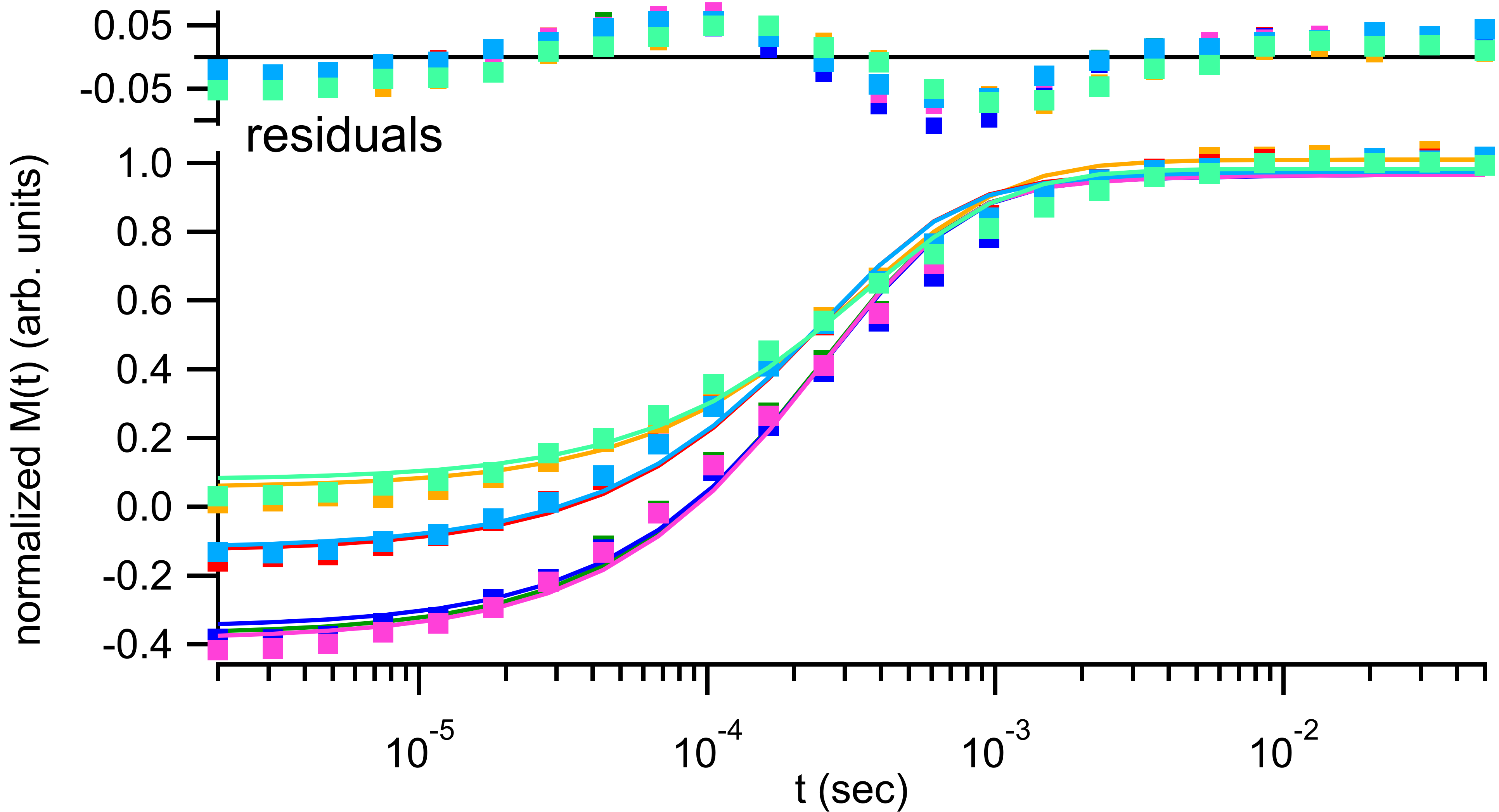}
  \caption{\label{fig:relaxfit}Magnetization recovery versus recovery time for the seven transitions as observed in Fig. \ref{fig:relaxation_channels}.  The solid lines are global fits to the relaxation of each transition with the constraint $W_{Q1}/W_{Q2}=4$. The residuals are plotted above and could be reduced by taking a distribution of relaxation rates into account.}
\end{figure}

The relaxation of a spin-7/2 nucleus in the presence of both quadrupolar and magnetic fluctuations is determined by the master equation:
\begin{equation}
    \frac{d\vec{p}}{dt}=\mathbb{W} \left(\vec{p} - \vec{p}_{EQ}\right),
    \label{eqn:master}
\end{equation}
where $p_m$ is the population of the $m^{th}$ nuclear spin energy level, $\vec{p}_{EQ}$ is the equilibrium population as determined by the Boltzmann distribution, and $\mathbb{W} = \mathbb{W}_m + \mathbb{W}_{Q1}+\mathbb{W}_{Q2}$, where:
\begin{widetext}
  \begin{equation}
    \mathbb{W}_m=W_m\left(
  \begin{array}{cccccccc}
    -7 & 7 & 0 & 0 & 0 & 0 & 0 & 0 \\
    7 & -19 & 12 & 0 & 0 & 0 & 0 & 0 \\
    0 & 12 & -27 & 15 & 0 & 0 & 0 & 0 \\
    0 & 0 & 15 & -31 & 16 & 0 & 0 & 0 \\
    0 & 0 & 0 & 16 & -31 & 15 & 0 & 0 \\
    0 & 0 & 0 & 0 & 15 & -27 & 12 & 0 \\
    0 & 0 & 0 & 0 & 0 & 12 & -19 & 7 \\
    0 & 0 & 0 & 0 & 0 & 0 & 7 & -7 \\
  \end{array}
\right),
\end{equation}
\begin{equation}
    \mathbb{W}_{Q1}=W_{Q1}\left(
   \begin{array}{cccccccc}
    -84 & 0 & 84 & 0 & 0 & 0 & 0 & 0 \\
    0 & -180 & 0 & 180 & 0 & 0 & 0 & 0 \\
    84 & 0 & -324 & 0 & 240 & 0 & 0 & 0 \\
    0 & 180 & 0 & -420 & 0 & 240 & 0 & 0 \\
    0 & 0 & 240 & 0 & -420 & 0 & 180 & 0 \\
    0 & 0 & 0 & 240 & 0 & -324 & 0 & 84 \\
    0 & 0 & 0 & 0 & 180 & 0 & -180 & 0  \\
    0 & 0 & 0 & 0 & 0 & 84 & 0 & -84 \\
  \end{array}
\right),
\end{equation}
and
\begin{equation}
    \mathbb{W}_{Q2}=W_{Q2}\left(
  \begin{array}{cccccccc}
    -252 & 252 & 0 & 0 & 0 & 0 & 0 & 0 \\
    252 & -444 & 192 & 0 & 0 & 0 & 0 & 0 \\
    0 & 192 & -252 & 60 & 0 & 0 & 0 & 0 \\
    0 & 0 & 60 & -60 & 0 & 0 & 0 & 0 \\
    0 & 0 & 0 & 0 & -60 & 60 & 0 & 0 \\
    0 & 0 & 0 & 0 & 60 & -252 & 192 & 0 \\
    0 & 0 & 0 & 0 & 0 & 192 & -444 & 252 \\
    0 & 0 & 0 & 0 & 0 & 0 & 252 & -252 \\
  \end{array}
\right).
\end{equation}
\end{widetext}
The initial conditions are set by the applied pulses, which in this case invert the populations of adjacent levels: $p_n(t=0) = p_{n-1, EQ}$, and $p_{n-1}(t=0) = p_{n, EQ}$ for the $n^{th}$ transition.
The time dependence of the magnetization is given by $p_{n}(t) - p_{n-1}(t)$, and consists of a weighted sum of normal mode exponential decays, in which the exponents and coefficients are determined by solution of \ref{eqn:master}.   There is no closed-form expression for the solution for the general case where $W_m, W_{Q1}, W_{Q2}>0$.  To fit our relaxation data, we solve Eq. \ref{eqn:master} numerically and minimize the global $\chi^2$ for all seven transitions ($-5/2\leq n \leq 7/2$) simultaneously, as illustrated in Fig. \ref{fig:relaxfit}. This process enables us to extract $W_m$, $W_{Q1}$ and $W_{Q2}$ independently. However, as can be already guessed from the opposite curvatures due to $W_{Q1}$ and $W_{Q2}$ in Fig. \ref{fig:relaxation_channels}a, in general there is a combination of large $W_{Q2}$ and small $W_{Q1}$ that has effectively no curvature and is thus difficult to distinguish from dominantly magnetic fluctuations. In \tmvo, for small fields and $T>T_Q$ unconstrained fits converge for small $W_m$ and $W_{Q2}/W_{Q2}\sim 10$ with a smaller $\chi^2$ value than our constrained fits. This can be expected since the unconstrained fit has more free parameters to fit to relaxation curves that are potentially stretched due to a distribution of relaxation rates.

\section{Field-Induced Enhancement of Quadrupolar Moment\label{sec:AppendixD}}

 Each Tm quadrupolar moment couples to the local lattice strain as:
\begin{equation}
    \mathcal{H}_{s} = \frac{1}{2} N\Omega c_{B2g}\varepsilon_{B2g}^2 + \sum_i V_s \varepsilon_{B2g} \hat{P}_{xy}(i)
\end{equation}
where $V_s$ is a coupling constant and $\varepsilon_{B2g} = \varepsilon_{xy}$ is the strain field.   Integrating out the lattice leads to the expression in Eq. \ref{eqn:exchange} with $J_{ij} = V_s^2/(2 c_{B2g}N\Omega)$ \cite{Gehring1975}. The  crystal field Hamiltonian of the Tm is given by:
\begin{equation}
\mathcal{H}_{CEF} = B_{2}^0 \hat{O}_2^0 + B_{4}^0 \hat{O}_4^0+B_{4}^4 \hat{O}_4^4+B_{6}^0 \hat{O}_6^0+B_{6}^4 \hat{O}_6^4
    \label{eqn:HCEF}
\end{equation}
where the $\hat{O}_l^m$ are the Steven's operators and the coefficients $B_l^m$ are well-known \cite{FisherNematicQCP,BleaneyTmVO4review}.  For temperatures well below the first excited crystal field state ($\sim 50$ K for \tmvo), the operator $\hat{P}_{xy}$ can be written as $E\hat\sigma_x$ in the ground state manifold.  However, an in-plane magnetic field will admix the excited crystal field states to the ground state wavefunctions to first order:
\begin{eqnarray}
    |\psi_{1}(B)\rangle &=& |\psi_{1}(0)\rangle + \sum_{i=3}^{13} \frac{g_J\mu_B B}{E_i-E_1} c_{1,i} |\psi_i(0)\rangle\\
    |\psi_{2}(B)\rangle &=& |\psi_{2}(0)\rangle + \sum_{i=3}^{13}\frac{g_J\mu_B B}{E_i-E_1} c_{2,i} |\psi_i(0)\rangle
\end{eqnarray}
where
\begin{equation}
    c_{(1,2),i} = \langle \psi_{i}(0)|(\hat{J}_x\cos(\phi) + \hat{J}_y\sin(\phi) ) |\psi_{1,2}(0)\rangle.
\end{equation}
Thus to lowest order in the ground state manifold $\hat{P}_{xy} = (E(0) + \xi B^2) \hat\sigma_x$, where $E(0) = 7.83$, and  $\xi = 0.059$ T$^{-2}$ for $\phi = 45^{\circ}$.

\end{document}